\documentclass[12pt]{article}
\usepackage{eurosym}
\usepackage{amssymb}
\usepackage{epsfig}
\usepackage{graphicx}
\usepackage{amsmath}
\usepackage{amsfonts}
\usepackage{amssymb}
\usepackage{float}
\usepackage{subfigure}
\usepackage{epstopdf}
\usepackage{float}
\DeclareGraphicsExtensions{.pdf} \textwidth = 16 truecm
\graphicspath{{Images/}}

\begin{document}

\title{\textbf{\Large Holographic conductivity of holographic
superconductors with higher order corrections}}
\author{\textbf{\ {\normalsize Ahmad Sheykhi$^{1,2}$\thanks{%
asheykhi@shirazu.ac.ir}}, {\normalsize Afsoon Ghazanfari$^{1}$},
{\normalsize Amin Dehyadegari$^{1}$}} \\
$^1$ {\normalsize Physics Department and Biruni Observatory, College of
Sciences, }\\
{\normalsize Shiraz University, Shiraz 71454, Iran}\\
$^2$ {\normalsize Research Institute for Astronomy and Astrophysics of
Maragha (RIAAM), }\\
{\normalsize P.O. Box 55134-441, Maragha, Iran}\\
[0.3cm] }
\date{}
\maketitle

\begin{abstract}
We analytically as well as numerically disclose the effects of the higher
order correction terms in the gravity and in the gauge field on the
properties of $s$-wave holographic superconductors. On the gravity side, we
consider the higher curvature Gauss-Bonnet corrections and on the gauge
field side, we add a quadratic correction term to the Maxwell Lagrangian. We
show that for this system, one can still obtain an analytical relation
between the critical temperature and the charge density. We also calculate
the critical exponent and the condensation value both analytically and
numerically. We use a variational method, based on the Sturm-Liouville
eigenvalue problem for our analytical study, as well as a numerical shooting
method in order to compare with our analytical results. For a fixed value of
the Gauss-Bonnet parameter, we observe that the critical temperature
decreases with increasing the nonlinearity of the gauge field. This implies
that the nonlinear correction term to the Maxwell electrodynamics make the
condensation harder. We also study the holographic conductivity of the
system and disclose the effects of Gauss-Bonnet and nonlinear parameters $%
\alpha$ and $b$ on the superconducting gap. We observe that for various
values of $\alpha $ and $b$, the real part of conductivity is proportional
to the frequency per temperature, $\omega /T$, as frequency is enough large.
Besides, the conductivity has a minimum in the imaginary part which is
shifted toward greater frequency with decreasing the temperature.
\end{abstract}

\section{Introduction}

The correspondence between the gravity in a $d$-dimensional anti-de Sitter
(AdS) spacetime and the conformal field theory (CFT) residing on the ($d-1$%
)-dimensional boundary of this spacetime, well-known as AdS/CFT
correspondence, provides an established method for calculating correlation
functions in a strongly interacting field theory using a dual classical
gravity description \cite{Mal}. It has been confirmed that this duality can
be applied for solving the problem of high temperature superconductors in
condensed matter physics \cite{Hartnol}. This is due to the fact the high
temperature superconductors are basically in a strong coupling regime, and
thus one expects that the holographic method could give some insights into
the pairing mechanism in these systems. Understanding the mechanism of high
temperatures superconductors has long been a mysteries problem in modern
condensed matter physics. Recently, it was suggested that it is logical to
understand the properties of high temperature superconductors on the
boundary of spacetime by considering a classical general relativity in one
higher dimensions. This idea is called the holographic superconductors (HSC)
\cite{Hartnol2,Har2,Hor} and has got a lot of attentions in the past decade.
According to the HSC proposal, in the gravity side, a Maxwell field and a
charged scalar field are introduced to describe the $U(1)$ symmetry and the
scalar operator in the dual field theory, respectively. This holographic
model undergoes a phase transition from black hole with no hair (normal
phase/conductor phase) to the case with scalar hair at low temperatures
(superconducting phase) \cite{Gub}.

Nowadays, the investigations on the HSC have attracted considerable
attention and become an active field of research. Let us review some works
in this direction. In the background of Schwarzschild AdS black holes in
Einstein gravity, the properties of HSC have been explored in \cite%
{Mus,RGC1,P.GWWY, P.BGRL, P.MRM, P.CW, P.ZGJZ,RGC2}. The studies were also
generalized to higher order gravity theories such as Gauss-Bonnet gravity
\cite{Wang1,Wang2,RGC3,Ruth,GBHSC}. It was argued that the critical
temperature of the HSC decreases with increasing the backreaction, although
the effect of the Gauss-Bonnet coupling is more subtle: the critical
temperature first decreases then increases as the coupling tends towards the
Chern-Simons value in a backreaction dependent fashion \cite{Ruth}. It was
confirmed that the critical exponent of the condensation in Gauss-Bonnet HSC
still obeys the mean field theory and has the value $1/2$ \cite{RGC3}. Other
studies on the holographic superconductor have been carried out in (see for
example \cite{Gr, RGC5,RGC6,RGC7,XHWang,Wang3,Wang4,Wang5,Wang6,Yao,JWPCh,
JJ1,JJ3,SG2,Shey2,Shey3,Shey4,SheyLN,SheyPM,a1,a2,a3,a4,a5,a6,a8} and
references therein).

It is also interesting to investigate the electrical conductivity of HSC in
the dual CFT as a function of frequency. In the AdS/CFT correspondence, the
electrical conductivity can be computed by looking at the linear response of
the system to fluctuations of the fields $A_{x}$ and $g_{tx}$ in the bulk.
These fluctuations are dual to the electric current $J_{x}$ and energy
current $T_{tx}$ operators in the CFT. In the context of linear Maxwell
field, the conductivity of HSC were computed in \cite{Hartnol,Har2,Hor}. In
the presence of nonlinear electrodynamics, the conductivity of HSC have been
investigated in \cite{logholg,powerholg}. Also, in the context of
Born-Infeld nonlinear electrodynamics, the optical properties of Lifshitz
HSC has been explored in \cite{opticalBI}. It was demonstrated that this
superconductor exhibits metamatrial property in low frequency of the
external electric field for certain region of nonlinear parameter. The
effects of the Weyle coupling parameter and Lifshitz dynamic exponent on the
conductivity of HSC have been explored in \cite{Mansoorimirza}. In Ref. \cite%
{BTZ}, a rotating BTZ black holes was considered as the gravity dual to $%
(1+1)$ dimensional superconductor. In this case, depending of the angular
momentum on the conductivity has been investigated. Recently, the authors of
\cite{analytic} have analytically computed the holographic conductivity of
HSC in the presence of Born-Infeld nonlinear electrodynamic by considering
the backreaction of the matter field on the bulk metric. Further
investigations on the holographic conductivity of HSC have been performed in
\cite{HLifshitz1,HLifshitz2}.

In this work, we will address the effects of the higher order corrections on
the holographic conductivity of the $s$-wave HSC. On the gravity side, we
will consider the Gauss-Bonnet curvature correction terms which is most
general action in the 5D spacetime and on the gauge field side we add the
quadratic nonlinear gauge term. We shall investigate the effects of these
correction terms on the imaginary and real parts of the electrical
conductivity of the system. With these correction terms, especially
including a Gauss-Bonnet correction to the 5D action, we have the most
general action with second-order field equations in 5D \cite{lovelock},
which provides the most general models for the $s$-wave HSC. Furthermore, in
an effective action approach to the string theory, the Gauss-Bonnet term
corresponds to the leading order quantum corrections to gravity, and its
presence guarantees a ghost-free action\cite{zwiebach}. The purpose of this
work is to anallytically as well as numerically explore the effects of these
correction terms on the properties of $s$-wave HSC.

The plan of the work is as follows. In section \ref{basic}, we will set up
our model of the HSC in Gauss-Bonnet gravity with nonlinear electrodynamics
in the probe limit and drive the equations of motion. In section \ref{Trho},
we analytically as well as numerically compute the relationship between the
critical temperature and the charge density of Gauss-Bonnet HSC. In section %
\ref{CExp}, we study condensation operator near the critical temperature
using analytical and numerical method. In section \ref{Cond} we investigate
the electrical conductivity of the HSC in Gauss Bonnet gravity with
nonlinear correction term to the Maxwell field. In particular, we shall find
the ratio of the gap frequency in conductivity to the critical temperature.
At last, we summarize and discuss our results in section \ref{Con}.

\section{HSC in Gauss-Bonnet gravity with nonlinear electrodynamics}

\label{basic} We consider the 5D Einstein-Gauss-Bonnet gravity in the
background of AdS spaces which is described by the action \cite{Gan},
\begin{equation}
S=\int d^{5}x\sqrt{-g}\left[ R-2\Lambda +\frac{\alpha }{2}\left(
R^{2}-4R^{\mu \nu }R_{\mu \nu }+R^{\mu \nu \rho \sigma }R_{\mu \nu \rho
\sigma }\right) +L_{M}\right] ,  \label{action}
\end{equation}%
where $\Lambda =-6/l^{2}$ is the cosmological constant of $5$-dimensional
AdS spacetime with radius $l$, $\alpha $ is the Gauss-Bonnet coefficient
with dimension $(\mathit{length})^2$, $R_{\mu \nu \rho \sigma}$, $R_{\mu
\nu} $ and $R$ are the Riemann curvature tensor, Ricci tensor, and the Ricci
scalar, respectively. For convenience, hereafter we set the AdS radius $l=1$%
. We consider the Lagrangian density of the matter field, $L_{M}$, as
\begin{equation}
L_{M}={L}_{NL}-|\nabla \psi -iqA\psi |^{2}-m^{2}|\psi |^{2}.  \label{L}
\end{equation}%
where $\psi $ is a scalar field, $q$ and $m$ are, respectively, the charge
and the mass of the scalar field, and the Lagrangian density of the
nonlinear electrodynamics is given by $\cite{HendiNL2,HendiNL3}$
\begin{equation}
\mathcal{L}_{\mathrm{NL}}=\mathcal{F}+b\mathcal{F}^{2}+O(\mathcal{F}^{4}),
\label{NL}
\end{equation}%
where $\mathcal{F}=-\frac{1}{4}F^{\mu \nu }F_{\mu \nu }$ is the Maxwell
Lagrangian and $b$ is a parameter. The term $b\mathcal{F}^{2}$ is the first
order leading nonlinear correction term to the Maxwell field. There are
several motivation for choosing the nonlinear Lagrangian in the form of (\ref%
{NL}). First, the series expansion of the three well-known Lagrangian of
nonlinear electrodynamics such as Born-Infeld, Logarithmic and Exponential
nonlinear electrodynamics have the form of (\ref{NL}) \cite{Hendi1}. Second,
calculating one-loop approximation of QED, it was shown \cite{Ritz} that the
effective Lagrangian is given by (\ref{NL}). Besides, if one neglect all
other gauge fields, one may arrive at the effective quadratic order of $U(1)$
as $\mathcal{F}^{2}$ \cite{Liu,Kats}. Furthermore, considering the next
order correction terms in the heterotic string effective action one can
obtain the $\mathcal{F}^{2}$ term as a corrections to the bosonic sector of
supergravity, which has the same order as the Gauss-Bonnet term \cite%
{Liu,Kats,An,Cai},
\begin{equation}
L_{\mathrm{cor}}=\beta \left[ \alpha \left( R^{2}-4R^{\mu \nu }R_{\mu \nu
}+R^{\mu \nu \rho \sigma }R_{\mu \nu \rho \sigma }\right) +b(F^{\mu \nu
}F_{\mu \nu })^{2}\right] .
\end{equation}%
The field equations can be obtained by varying action (\ref{action}) with
respect to the metric $g_{\mu \nu}$, the scalar field $\psi$, and and the
gauge field $A_{\mu}$. We find
\begin{eqnarray}  \label{FE1}
&&R_{\mu \nu }-\frac{(R-2\Lambda )}{2}g_{\mu \nu }-\frac{\alpha }{2}\left\{
\frac{1}{2}g_{\mu \nu }\left( R^{2}-4R^{\rho \sigma }R_{\rho \sigma
}+R^{\kappa \lambda \rho \sigma }R_{\kappa \lambda \rho \sigma }\right)
\right.  \notag \\
&& \left.-2RR_{\mu \nu }+4R_{\mu \lambda }R_{\ \nu }^{\lambda }+4R_{\mu \rho
\nu \sigma }R^{\rho \sigma }-2R_{\mu }^{ \ \rho \sigma \lambda }R_{\nu \rho
\sigma \lambda }\right\}=T_{\mu \nu },
\end{eqnarray}%
\begin{equation}  \label{FE2}
(\nabla _{\mu }-iqA_{\mu })(\nabla ^{\mu }-iqA^{\mu })\psi -m^{2}\psi =0,
\end{equation}%
\begin{equation}
\nabla_{\mu }\left[ \left( 1+2b\mathcal{F}\right) F^{\mu \nu }\right] =iq%
\left[ \psi ^{\ast }(\nabla ^{\nu }-iqA^{\nu })\psi -\psi (\nabla ^{\nu
}+iqA^{\nu })\psi ^{\ast }\right],  \label{FE3}
\end{equation}%
{where }$T_{\mu \nu }$ is the matter-stress tensor
\begin{eqnarray}
&&T_{\mu \nu }=\frac{1}{2}\left( \mathcal{F}+b\mathcal{F}^{2}\right) g_{\mu
\nu }-2\left( 1+2b\mathcal{F}\right) F_{\mu \rho }F_{\nu }^{\ \rho }-\frac{1%
}{2}m^{2}|\psi |^{2}g_{\mu \nu }-\frac{1}{2}g_{\mu \nu }|\nabla \psi
-iqA\psi |^{2}  \notag \\
&&+\frac{1}{2}\left[ (\nabla ^{\nu }-iqA^{\nu })\psi (\nabla ^{\mu
}+iqA^{\mu })\psi^{\ast } +\mu \leftrightarrow \nu \right] .
\end{eqnarray}%
The metric of a planar Schwarzschild-AdS black hole in 5D is \cite{CaiGB}
\begin{equation}  \label{metr}
ds^{2}=-f(r)dt^{2}+\frac{dr^{2}}{f(r)}+r^{2}(dx^{2}+dy^{2}+dz^{2}),
\end{equation}%
with
\begin{equation}
f(r)=\frac{r^{2}}{2\alpha }\left( 1-\sqrt{1-4\alpha \left( 1-\frac{r_{+}^{4}%
}{r^{4}}\right) }\right) .
\end{equation}%
The Hawking temperature at the horizon can be written in the form
\begin{equation}
T=\frac{f^{^{\prime }}(r)}{4\pi }=\frac{r_{+}}{\pi }.  \label{temp}
\end{equation}%
It is worthwhile to note that in the limit $r\rightarrow \infty $, we can
obtain
\begin{equation}
f(r)\sim \frac{r^{2}}{2\alpha }\left[ 1-\sqrt{1-4\alpha }\right] ,  \label{f}
\end{equation}%
so we can introduce the effective AdS radius as
\begin{equation}
L_{\mathrm{eff}}^{2}=\frac{2\alpha }{1-\sqrt{1-4\alpha }}.
\end{equation}%
We choose the gauge and the scalar fields in the form \cite{Hartnol}
\begin{equation}  \label{scalar2}
A_{\mu }=(\phi (r),0,0,0,0),~\ \ \ ~~~~\psi =\psi (r),
\end{equation}
Inserting the metric (\ref{metr}) and the gauge and scalar fields (\ref%
{scalar2}) in the field equations (\ref{FE2}) and (\ref{FE3}), we arrive at
\begin{equation}
\partial _{r}^{2}\psi +\left( \frac{3}{r}+\frac{\partial _{r}f}{f}\right)
\partial _{r}\psi +\left( \frac{\phi ^{2}}{f^{2}}-\frac{m^{2}}{f}\right)
\psi =0,  \label{eqpsi1}
\end{equation}%
\begin{equation}
\partial _{r}^{2}\phi +\frac{3}{r}\left( 1-2b(\partial _{r}\phi )^{2}\right)
\partial _{r}\phi -\frac{2\psi ^{2}\phi }{f}\left( 1-3b(\partial _{r}\phi
)^{2}\right) =0.  \label{eqphi1}
\end{equation}%
The horizon radius is defined as the root of $f(r_{+})=0$. The regularity
condition for the gauge field $A_{t}$ on the horizon $r_{+}$, implies the
boundary condition $\phi(r_+)=0$, which substituting in Eq. (\ref{eqpsi1})
yields
\begin{equation}
\ ~~~~~\psi(r_+)=\frac{\partial_{r}f(r_+)}{m^{2}}\partial_{r}\psi(r_+).
\end{equation}
Near the AdS boundary ($r\rightarrow\infty $) the asymptotic behaviors of
the solutions are given by
\begin{eqnarray}
\phi(r)&=&\mu - \frac{\rho}{r^{2}} ,  \label{phi} \\
\psi(r)&=&\frac{\psi_{-}}{r^{\Delta_{-}}}+\frac{\psi_{+}}{r^{\Delta_{+}}},
\end{eqnarray}
where $\Delta_{\pm} =2\pm\sqrt{4+m^{2}L_{\mathrm{eff}}^{2}}$. It is clear
that we should have $m^{2}L_{\mathrm{eff}}^{2}\geq-4$. The value of $%
\Delta_{\pm}$ depend on $\tilde{m}^2=m^{2}L_{\mathrm{eff}}^{2}$. For
example, setting $\tilde{m}^2=-3$, we have $\Delta_+=3$ and $\Delta_-=1$.
The coefficients $\psi_{\pm}$ correspond to the vacuum expectation values of
the condensate operator, namely $\psi_{\pm}=<\mathcal{O_{\pm}}>$, where $%
\mathcal{O_{\pm}}$ is the dual operator to the scalar field with the
conformal dimension $\Delta_{\pm}$. Following \cite{Hartnol}, we can impose
the boundary condition in which either $\psi_-$ or $\psi_+$ vanishes, so
that the theory is stable in the asymptotic AdS region. In what follow, we
set $\psi_{-}=0 $ and take $\psi_{+}=\langle \mathcal{O_{+}} \rangle $ non
zero. The interpretation of the parameters $\mu $ and $\rho $, also comes
from the gauge/gravity dictionary and are, respectively, interpreted as the
chemical potential and charge density of the conformal field theory on the
boundary.

\section{Relation between critical temperature and charge density}

\label{Trho}

In this section, we are going to study the critical temperature of HSC, when
the higher order corrections to the gravity side as well as the gauge field
is taken into account. We shall continue our study both analytically and
numerically and compare the two method with each other.

\subsection{Analytical method}

Here, we analytically obtain the relation between the critical temperature
and the charge density of Gauss-Bonnet HSC. To do this, we first transform
the coordinate $r$ to $z$, such that $z={r_{+}}/{r}$. Using this new
coordinates, the equations of motion (\ref{eqpsi1}) and (\ref{eqphi1}) can
be rewritten as
\begin{equation}
\partial _{z}^{2}\phi -\frac{1}{z}\partial _{z}\phi +\frac{6bz^{3}}{r_{+}^{2}%
}(\partial _{z}\phi )^{3}-\frac{2\psi ^{2}\phi r_{+}^{2}}{fz^{4}}+\frac{%
6b(\partial _{z}\phi )^{2}\phi \psi ^{2}}{f}=0,  \label{eq4}
\end{equation}%
\begin{equation}
\partial _{z}^{2}\psi +(\frac{\partial _{z}f}{f} -\frac{1}{z}) \partial
_{z}\psi +\frac{r_{+}^{2}}{z^{4}}(\frac{\phi ^{2}}{f^{2}}-\frac{m^{2}}{f}%
)\psi=0.  \label{eq5}
\end{equation}%
Near the critical temperature ($T=T_{c}$) we have $\psi =0$, and thus
equation (\ref{eq4}) reduces to
\begin{equation}
\partial _{z}^{2}\phi -\frac{1}{z}\partial _{z}\phi +\frac{6bz^{3}}{r_{+}^{2}%
}(\partial _{z}\phi )^{3}=0.
\end{equation}%
Solving the above equation for the small value of nonlinear parameter $b$,
we find
\begin{equation}
\phi (z)=\lambda r_{+c}(1-z^{2})\left[ 1-\frac{b\lambda ^{2}}{2}\xi (z)%
\right] +O(b^{2}),  \label{eq6}
\end{equation}%
where
\begin{equation}
\xi (z)=(1+z^{2})(1+z^{4}),~~~\ \ \ ~~~\lambda =\frac{\rho }{r_{+c}^{3}}.
\label{lambda}
\end{equation}%
Next, we consider the boundary conditions for $\psi $ near the critical
point ($T\rightarrow T_{c}$). We assume $\psi $ has the following form \cite%
{Gan2}
\begin{equation}
\psi |_{z\rightarrow 0}\sim \frac{\langle \mathcal{O_{+}}\rangle }{r_{+}^{3}}%
z^{3}F(z),  \label{eq7}
\end{equation}%
where $F(z)$ is the trial function near the boundary $z=0$, which satisfies
the boundary conditions $F(0)=1$ and $F^{\prime }(0)=0$. Substituting Eqs. (%
\ref{eq6}) and (\ref{eq7}) in Eq. (\ref{eq5}) one arrives at
\begin{equation}
F^{^{\prime \prime }}+p(z)F^{^{\prime }}+q(z)F+\lambda ^{2}w(z)F=0,
\label{eq8}
\end{equation}%
where the prime now indicates the derivative with respect to $z$, and $p(z)$%
, $q(z)$ and $w(z)$ read
\begin{eqnarray}
p(z) &=&\frac{3(1-\sqrt{1-4\alpha +4\alpha z^{4}})-12\alpha +20\alpha z^{4}}{%
z[1-4\alpha +4\alpha z^{4}-\sqrt{1-4\alpha +4\alpha z^{4}}]}, \\
q(z) &=&\frac{1}{z^{2}}\left[ \frac{3(1-4\alpha -4\alpha z^{4}-\sqrt{%
1-4\alpha +4\alpha z^{4}})}{\sqrt{1-4\alpha +4\alpha z^{4}}-1+4\alpha
-4\alpha z^{4}}+\frac{2m^{2}\alpha }{\sqrt{1-4\alpha +4\alpha z^{4}}-1}%
\right] , \\
w(z) &=&\frac{4\alpha ^{2}(1-z^{2})^{2}(1-\frac{b}{2}\lambda ^{2}\xi (z))^{2}%
}{(1-\sqrt{1-4\alpha +4\alpha z^{4}})^{2}}.
\end{eqnarray}%
It is a matter of calculations to convert Eq. (\ref{eq8}) to the standard
form of the Sturm-Liouville equation
\begin{equation}
(T(z)F^{^{\prime }}(z))^{^{\prime }}-Q(z)F(z)+\lambda ^{2}P(z)F(z)=0,
\label{SL}
\end{equation}%
where,
\begin{eqnarray}
T(z) &=&\frac{z^{3}}{2\sqrt{\alpha }}(\sqrt{1-4\alpha +4\alpha z^{4}}%
-1)\approx z^{3}\sqrt{\alpha }(z^{4}-1)[1-\alpha (z^{4}-1)], \\
Q(z) &=&-T(z)q(z)\approx -3z\sqrt{\alpha }(3z^{4}+6\alpha z^{4}-7\alpha
z^{8}), \\
P(z) &=&T(z)w(z)\approx \frac{\sqrt{\alpha }z^{3}(z^{2}-1)(1+\alpha
(z^{4}-1))(1-\frac{b}{2}\lambda ^{2}\xi (z))^{2}}{z^{2}+1}.  \label{eq9}
\end{eqnarray}%
In the above equations we have only kept the terms up to order $\alpha
^{3/2} $. Next, we perform a perturbative expansion $b\lambda ^{2}$ and
retain only the terms that are linear in $b$ such that
\begin{equation}
b\lambda ^{2}=b(\lambda ^{2}|_{b=0})+O(b^{2}),
\end{equation}%
where $\lambda ^{2}|_{b=0}$ is the value of $\lambda ^{2}$ for $b=0$. Thus
we can rewrite Eq. (\ref{eq9}) as
\begin{equation}
P(z)\approx \frac{\sqrt{\alpha }z^{3}(z^{2}-1)(1+\alpha
(z^{4}-1))(1-b(\lambda ^{2}|_{b=0})\xi (z))}{z^{2}+1}.
\end{equation}%
Employing the Sturm-Liouville eigenvalues problem, the eigenvalues of Eq. (%
\ref{SL}) can be obtained by varying the following function
\begin{equation}
\lambda ^{2}=\frac{\int_{0}^{1}dz(T(z)(F^{^{\prime }}(z))^{2}+Q(z)F^{2}(z))}{%
\int_{0}^{1}dzP(z)F^{2}(z)},  \label{SL2}
\end{equation}%
where we also choose $F(z)=1-az^{2}$ and $m^{2}=-3/L_{\mathrm{eff}}^{2}$ to
appraise this expression. At last, using Eqs. (\ref{temp}) and (\ref{lambda}%
), for $T\sim T_{c}$, one can obtain
\begin{equation}
T_{c}=\zeta \rho ^{1/3},  \label{Tc}
\end{equation}%
where $\zeta =\frac{1}{\pi \lambda _{\mathrm{min}}^{1/3}}$ and $\lambda _{%
\mathrm{min}}$ is the minimum eigenvalue which can be obtained by variation
of Eq. (\ref{SL2}). Our strategy, in the analytical method, for calculating
the critical temperature for condensation is to minimize the function (\ref%
{SL2}) with respect to the coefficient $a$ by fixing other parameters of the
model such as $b$ and $\alpha $. Then, we obtain $\lambda _{\mathrm{min}}$
and hence the maximum value of $T_{c}/\rho ^{1/3}$ can be deduced through
relation (\ref{Tc}). {As an example, we bring the details of our calculation
for $\alpha =0.01$ and $b=0.01$. In this case Eq. (\ref{SL2}) reduces to
\begin{equation}
\lambda ^{2}=\frac{0.150900-0.226000a+0.117119a^{2}}{%
0.003844-0.003350a+0.000911a^{2}},
\end{equation}%
whose minimum is $\lambda_{\mathrm{min}}=25.6427$ at $a =0.747087$. And thus
according to Eq. (\ref{Tc}), the critical temperature becomes }$%
T_{c}=0.185363\rho ^{1/3}$\textbf{.} In tables 1, 2 and 3, we summarize our
results for $\lambda _{\mathrm{min}}$ and $\zeta $ for different values of
the parameters $\alpha $, $a$ and $b$. This table shows that for a small and
fixed value of $\alpha $, with increasing the nonlinear parameter $b$, the
value of $\zeta =T_{c}/\rho ^{1/3}$ decreases as well. As we shall see in
the next section this results is in a very good agreement with the numerical
results.

\subsection{Numerical method}

Now, we numerically investigate the critical behavior of the HSC in
Gauss-Bonnet gravity with quadratic correction term to the gauge field. For
the numerical study we employ the shooting method \cite{Shoot}. For
simplicity we assume $r_{+}=1$, and thus Eqs. (\ref{eq4}) and (\ref{eq5})for
$\phi $ and $\psi $ reduces to

\begin{equation}
\partial _{z}^{2}\phi -\frac{1}{z}\partial _{z}\phi +6bz^{3}(\partial
_{z}\phi )^{3}-\frac{2\psi ^{2}\phi }{fz^{4}}+\frac{6b(\partial _{z}\phi
)^{2}\phi \psi ^{2}}{f}=0,  \label{EQPHI}
\end{equation}
\begin{equation}
\partial _{z}^{2}\psi +\left( \frac{\partial _{z}f}{f}-\frac{1}{z}\right)
\partial _{z}\psi +\frac{1}{z^{4}}\left( \frac{\phi ^{2}}{f^{2}}-\frac{m^{2}%
}{f}\right) \psi =0.  \label{EQSY}
\end{equation}%
Near the horizon ($z=1$), we can expand $\phi $ and $\psi $ as

\begin{equation}
\phi \approx \phi ^{^{\prime }}(1)(1-z)+\frac{\phi ^{^{\prime \prime }}(1)}{2%
}(1-z)^{2}+\dots ,
\end{equation}%
\begin{equation}
\psi \approx \psi (1)+\psi ^{^{\prime }}(1)(1-z)+\frac{\psi ^{^{\prime
\prime }}(1)}{2}(1-z)^{2}+\dots ,
\end{equation}%
while near the AdS boundary ($z=0$), they behave like

\begin{equation}
\phi \approx \mu -\rho z^{2},
\end{equation}
\begin{equation}
\psi \approx \psi _{-}z^{\Delta _{-}}+\psi _{+}z^{\Delta _{-}}.
\end{equation}%
%
%
%
%
%
%
%
%
%
%
%
%
%
%
%
%
%
%
%
%
%
%
%
%
%
%
%
%
%
%
%
%
%
%
%
%
%
%
\begin{figure*}[t]
\begin{center}
\begin{minipage}[b]{0.325\textwidth}\begin{center}
       \subfigure[~$\alpha=0.0001,b=0.01$]{
                \label{fig1a}\includegraphics[width=\textwidth]{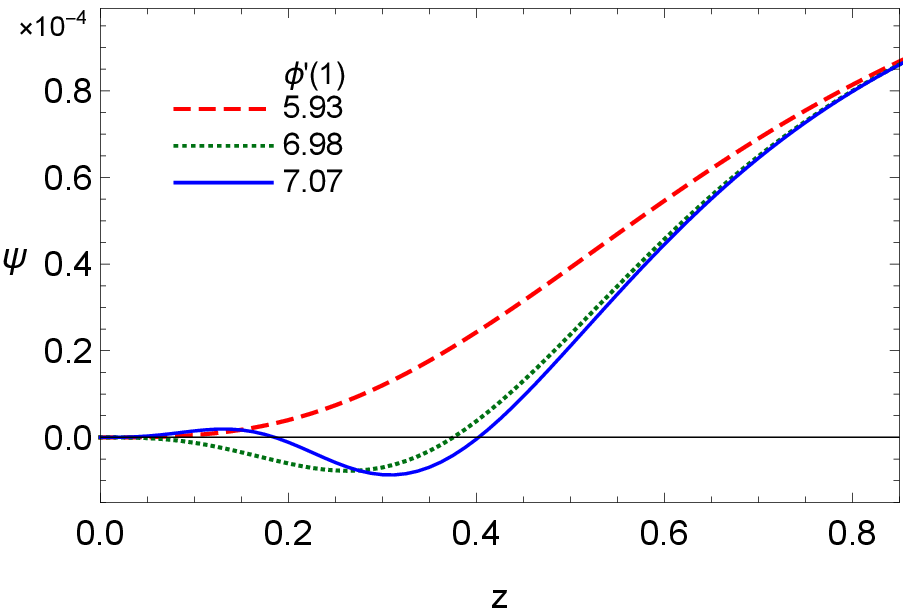}\qquad}
    \end{center}\end{minipage} \hskip+0cm
\begin{minipage}[b]{0.325\textwidth}\begin{center}
        \subfigure[~$\alpha=0.01,b=0$]{
                 \label{fig1b}\includegraphics[width=\textwidth]{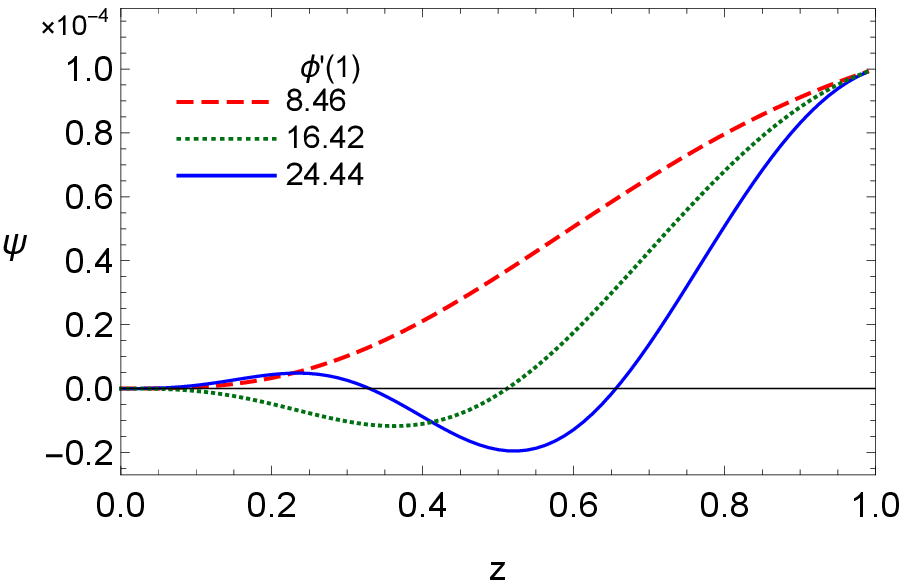}\qquad}
    \end{center}\end{minipage} \hskip0cm
\begin{minipage}[b]{0.325\textwidth}\begin{center}
         \subfigure[~$\alpha=0.01,b=0.01$]{
                  \label{fig1c}\includegraphics[width=\textwidth]{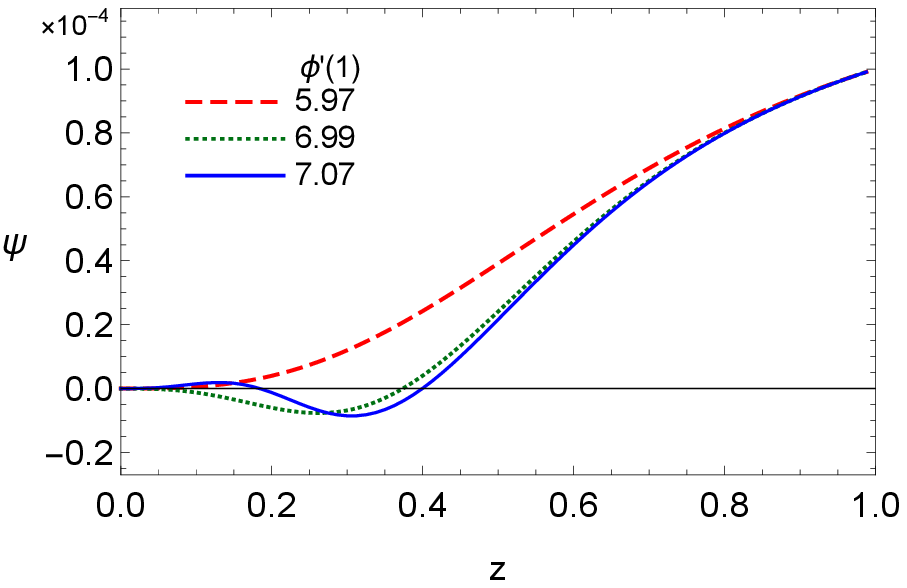}\qquad}
    \end{center}\end{minipage} \hskip0cm
\begin{minipage}[b]{0.325\textwidth}\begin{center}
             \subfigure[~$\alpha=0.1,b=0$]{
                      \label{fig1d}\includegraphics[width=\textwidth]{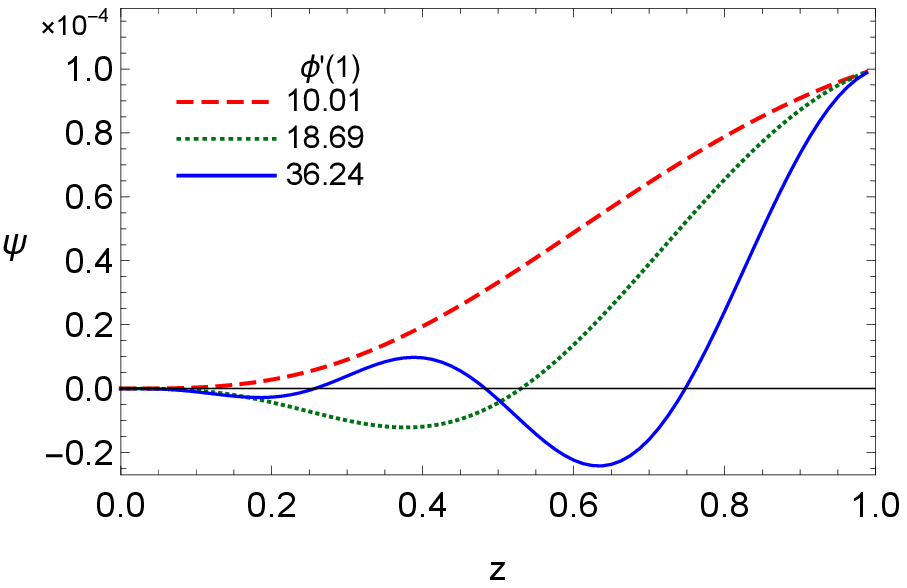}\qquad}
        \end{center}\end{minipage} \hskip0cm
\end{center}
\caption{The behavior of $\protect\psi (z)$ versus $z$ for Gauss-Bonnet HSC
and for $\tilde{m}^{2}=-3$ and different $\protect\alpha $ and $b$.}
\label{fig1}
\end{figure*}
\begin{table}[tbh]
\label{table1} \centering
\begin{tabular}{ccccccc}
\hline\hline
$b$ & $a $ & $\lambda_{\mathrm{min}}^{2} $ & $\zeta_{SL}\left(=\frac{1}{\pi
\lambda_{\mathrm{min}}^{1/3}}\right)$ & $\zeta_{\mathrm{Numerical}}$ &  &  \\%
[0.05ex] \hline
0 & 0.721772 & 18.2331 & 0.196204 & 0.197957 &  &  \\
0.01 & 0.747560 & 25.0682 & 0.186064 & 0.181057 &  &  \\
0.02 & 0.777479 & 36.1392 & 0.175060 & 0.165642 &  &  \\
0.03 & 0.809805 & 55.2025 & 0.163126 & 0.151671 &  &  \\[0.5ex] \hline
\end{tabular}%
\caption{Comparison of analytical and numerical values of $\protect\zeta %
=T_{c}/\protect\rho ^{1/3}$ for $\protect\alpha =0.0001$.}
\end{table}

\begin{table}[tbh]
\label{table2} \centering
\begin{tabular}{ccccccc}
\hline\hline
$b$ & $a $ & $\lambda_{\mathrm{min}}^{2} $ & $\zeta_{SL}\left(=\frac{1}{\pi
\lambda_{\mathrm{min}}^{1/3}}\right)$ & $\zeta_{\mathrm{Numerical}}$ &  &  \\%
[0.05ex] \hline
0 & 0.720561 & 18.5392 & 0.195660 & 0.196843 &  &  \\
0.01 & 0.747087 & 25.6427 & 0.185363 & 0.179433 &  &  \\
0.02 & 0.777900 & 37.2577 & 0.174173 & 0.163592 &  &  \\
0.03 & 0.811046 & 57.4907 & 0.162026 & 0.149279 &  &  \\[0.5ex] \hline
\end{tabular}%
\caption{Comparison of analytical and numerical values of $\protect\zeta %
=T_{c}/\protect\rho ^{1/3}$ for $\protect\alpha =0.01$.}
\end{table}
\begin{table}[tbh]
\label{table3} \centering
\begin{tabular}{ccccccc}
\hline\hline
$b$ & $a $ & $\lambda_{\mathrm{min}}^{2} $ & $\zeta_{SL}\left(=\frac{1}{\pi
\lambda_{\mathrm{min}}^{1/3}}\right)$ & $\zeta_{\mathrm{Numerical}}$ &  &  \\%
[0.05ex] \hline
0 & 0.709061 & 21.5679 & 0.190787 & 0.186114 &  &  \\
0.01 & 0.743348 & 31.6982 & 0.178928 & 0.162657 &  &  \\
0.02 & 0.783517 & 49.9342 & 0.165876 & 0.141983 &  &  \\
0.03 & 0.823574 & 85.6808 & 0.151601 & 0.124005 &  &  \\[0.5ex] \hline
\end{tabular}%
\caption{Comparison of analytical and numerical values of $\protect\zeta %
=T_{c}/\protect\rho ^{1/3}$ for $\protect\alpha =0.1$.}
\end{table}
We calculate $\phi ^{^{\prime \prime }}(1)$, $\psi ^{^{\prime }}(1)$ and $%
\psi ^{^{\prime \prime }}(1)$ in the term of $\psi (1)$ and $\phi ^{^{\prime
}}(1)$ by using the equations of motion for $\phi $ and $\psi $, namely Eqs.
(\ref{EQPHI}) and (\ref{EQSY}), respectively. Since near the critical point $%
\psi $ is very small, thus we choose $\psi (1)=0.0001$. Our strategy for
using the shooting method is as follows. For specific value of the reduced
scalar field mass $\tilde{m}^{2}$, we can perform numerical calculation near
the horizon boundary with one shooting parameter $\phi ^{^{\prime }}(1)$ to
get proper solutions at the infinite boundary. For specific values of $\phi
^{^{\prime }}(1)$, we impose the boundary condition $\psi _{-}=0$. We also
calculate the analytical values and numerical values of $\zeta $ for
different $b$. We compare our numerical results with analytical in tables $1$%
, $2$ and $3$.

In Fig. \ref{fig1}, we plot $\psi $ versus $z$ for three first boundary
condition $\phi ^{^{\prime }}(1)$, $\tilde{m}^{2}=-3$ and different values
of Gauss-Bonnet coefficient $\alpha $ and nonlinear parameter $b$. In the
absence of quadratic correction term ($b=0$), our results exactly coincide
with those presented in \cite{RGC3,Gr}. The acceptable diagram for us is the
red one in each plot since there is nothing in the bulk to effect on speed
of the wave, so the diagram of $\psi $ will be stable. From tables $1-3$, it
is evident that when $b$ becomes larger the condensation gets harder.
Similar behavior can be seen for the fixed value of $b$ and different values
of $\alpha $, namely the critical temperature reduces and condensation
becomes harder when the Gauss-Bonnet coupling parameter $\alpha $ gets
larger.

\section{Critical exponent and condensation values}

\label{CExp} In this section, our aim is to calculate the critical exponent
of HSC with first order correction terms in gravity and gauge field. Again,
we continue our studying both analytically and numerically.

\subsection{Analytical method}

We would like to obtain the critical exponent and the condensation values of
the condensation operator near the critical temperature using the analytical
method. Inserting Eq. (\ref{eq7}) into Eq. (\ref{eq4}), we get
\begin{equation}
\partial _{z}^{2}\phi -\frac{1}{z}\partial _{z}\phi +\frac{6bz^{3}}{r_{+}^{2}%
}(\partial _{z}\phi )^{3}=\frac{\langle \mathcal{O_{+}} \rangle^{2}}{%
r_{+}^{4}}\mathcal{B}\phi ,  \label{eq10}
\end{equation}%
and,
\begin{equation}
\mathcal{B}=\frac{2z^{2}}{f}\left( 1-\frac{3bz^{4}(\partial _{z}\phi )^{2}}{%
r_{+}^{2}}\right) F^{2}(z).
\end{equation}%
Near the critical temperature, $\frac{\langle \mathcal{O_{+}} \rangle^{2}}{%
r_{+}^{4}}$ is a very small and thus we can expand $\phi (z)$ as
\begin{equation}
\frac{\phi (z)}{r_{+}}=\lambda (1-z^{2})\left[ 1-\frac{b\lambda ^{2}}{2}\xi
(z)\right] +\frac{\langle \mathcal{O_{+}} \rangle^{2}}{r_{+}^{4}}\chi (z).
\label{eq11}
\end{equation}
where $\chi$ satisfies the following boundary condition
\begin{eqnarray}  \label{chi}
\chi (1)=\chi ^{^{\prime }}(1)=0.
\end{eqnarray}
With the help of Eq. (\ref{eq11}), and comparing the coefficient of $\frac{%
\langle \mathcal{O_{+}} \rangle^{2}}{r_{+}^{4}}$ on both sides, Eq. (\ref%
{eq10}) leads to
\begin{equation}
\chi ^{^{\prime \prime }}(z)-\frac{\chi ^{^{\prime }}}{z}+72b\lambda
^{2}z^{5}\chi ^{^{\prime }}=\lambda \mathcal{B}(1-z^{2})\left( 1-\frac{b}{2}%
\lambda ^{2}\xi (z)\right) .  \label{eq12}
\end{equation}%
From Eq. (\ref{eq12}) we figure out, in the limit $z\rightarrow 0$, the
following equation
\begin{equation}
\chi ^{^{\prime \prime }}(0)=\frac{\chi ^{^{\prime }}(z)}{z}|_{z\rightarrow
0}.  \label{eq13}
\end{equation}%
It is a matter of calculations to shaw that Eq. (\ref{eq12}) can be written
\begin{equation}
\frac{d}{dz}\left( e^{12b\lambda ^{2}z^{6}}\frac{\chi ^{^{\prime }}}{z}%
\right) =\lambda \frac{2z^{3}}{r_{+}^{2}}\frac{e^{12b\lambda ^{2}z^{6}}(1-%
\frac{b}{2}\lambda ^{2}\Gamma (z))}{(1+z^{2})(1+\alpha (1-z^{4}))}F^{2}(z),
\label{eq14}
\end{equation}%
where $\Gamma (z)=1+z^{2}+z^{4}+25z^{6}$. Integrating both sides of the
above equation in the interval $[0,1]$ and using the boundary condition (\ref%
{chi}), we arrive at
\begin{eqnarray}
\frac{\chi ^{^{\prime }}}{z}|_{z\rightarrow 0}=-\frac{\lambda }{r_{+}^{2}}%
\mathcal{A},  \label{eq15}
\end{eqnarray}%
where
\begin{eqnarray}  \label{A}
\mathcal{A} &\approx &\int_{0}^{1}\frac{2z^{3}F^{2}(z)\left[1-\frac{b}{2}%
\lambda ^{2}(1+z^{2}+z^{4}+z^{6})\right][1-\alpha (1-z^{4})]}{1+z^{2}}dz.
\notag
\end{eqnarray}%
Combining Eqs. (\ref{phi}) and (\ref{eq11}), we achieve
\begin{equation}  \label{eq16}
\frac{\mu}{r_{+}} -\frac{\rho }{r_{+}^{3}}z^{2}=\lambda (1-z^{2})\left[ 1-%
\frac{b\lambda ^{2}}{2}\xi (z)\right] +\frac{\langle \mathcal{O_{+}}
\rangle^{2}}{r_{+}^{4}}(\chi (0)+z\chi ^{^{\prime }}(0)+\frac{z^{2}}{2}\chi
^{^{\prime \prime }}(0)+...),
\end{equation}
where in the last step we have expanded $\chi(z)$ around $z = 0$. Equating
the coefficients of $z^2$ on both sides of Eq. (\ref{eq16}), we find
\begin{eqnarray}  \label{RL}
\frac{\rho}{r_{+} ^3}= \lambda \left(1+\frac{\langle \mathcal{O_{+}} \rangle
^2}{r_{+} ^6} \mathcal{A} \right).
\end{eqnarray}
Using the fact that $\lambda={\rho}/{r^3_{+c}}$ as well as definition (\ref%
{temp}), we can obtain the order parameter $\langle \mathcal{O_{+}} \rangle$
near the critical temperature $T_{c}$ as

\begin{eqnarray}
\langle \mathcal{O_{+}} \rangle =\gamma \pi ^{3}T_{c}^{3}\sqrt{1-\frac{T}{%
T_{c}}},  \label{gamma}
\end{eqnarray}%
where
\begin{eqnarray}
\gamma =\sqrt{\frac{6}{\mathcal{A}}}.
\end{eqnarray}
From Eq. (\ref{gamma}) we observe that the critical exponent has the mean
field value $1/2$, which is independent of the nonlinear parameter $b$ and
Gauss-Bonnet parameter $\alpha$. It is worth noting that $\langle \mathcal{O}
\rangle$ is zero at $T = T_{c}$ and condensation occurs for $T < T_{c}$. We
shall back to calculation the condensation value $\gamma $ in the next
subsection.
\begin{figure*}[t]
\begin{center}
\begin{minipage}[b]{0.325\textwidth}\begin{center}
       \subfigure[~$\alpha=0.0001$]{
                \label{fig2a}\includegraphics[width=\textwidth]{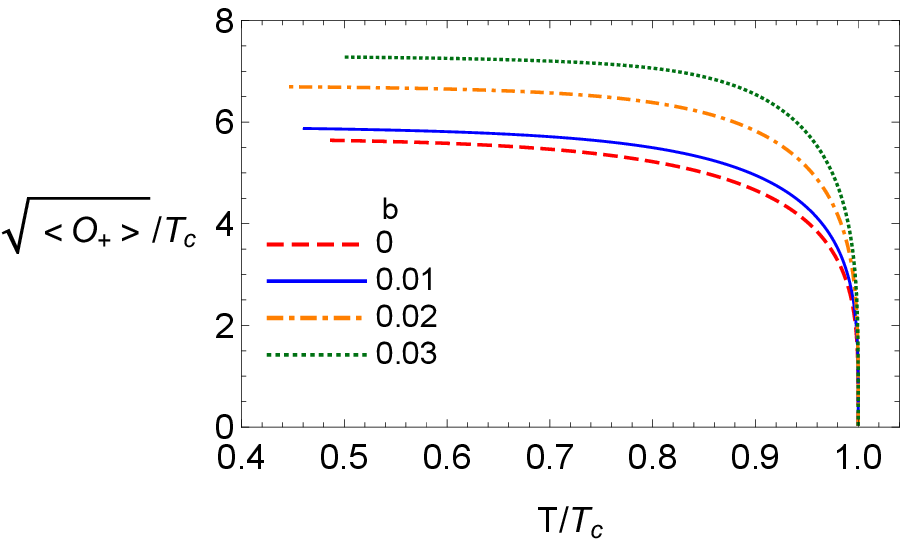}\qquad}
    \end{center}\end{minipage} \hskip+0cm
\begin{minipage}[b]{0.325\textwidth}\begin{center}
        \subfigure[~$\alpha=0.01$]{
                 \label{fig2b}\includegraphics[width=\textwidth]{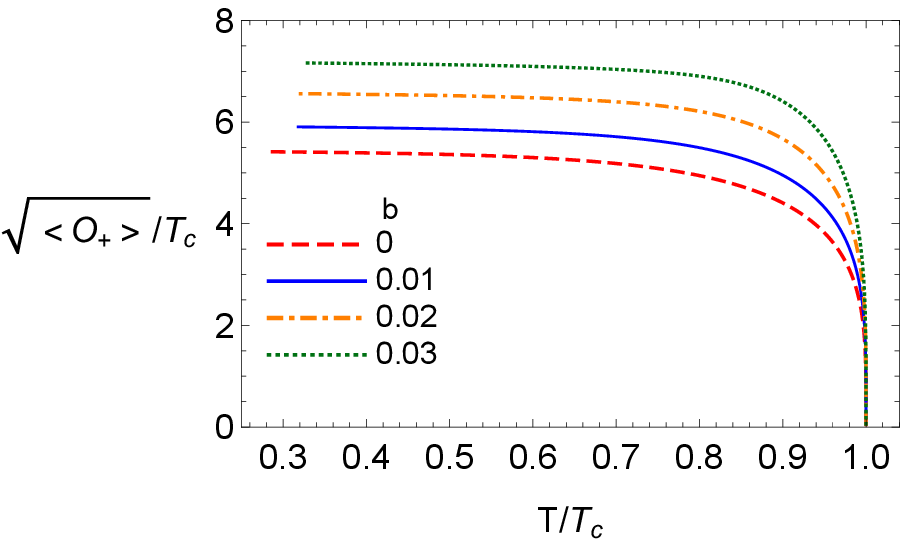}\qquad}
    \end{center}\end{minipage} \hskip0cm
\begin{minipage}[b]{0.325\textwidth}\begin{center}
         \subfigure[~$\alpha=0.1$]{
                  \label{fig2c}\includegraphics[width=\textwidth]{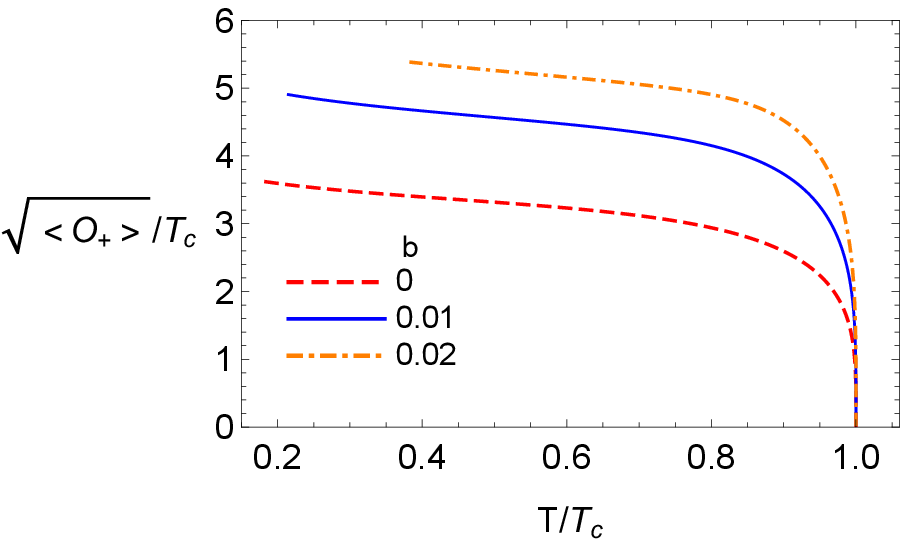}\qquad}
    \end{center}\end{minipage} \hskip0cm
\end{center}
\caption{The condensate operator $<\mathcal{O} _{+}>$ as a function of
temperature for different values of $b$ and various values of $\protect%
\alpha $, where we have set $<\mathcal{O} _{-}>=0$ and $\tilde{m}^2=-3$.}
\label{fig2}
\end{figure*}

\subsection{Numerical method}

We use the numerical method to explore the behaviour of the condensate
operator $\langle\mathcal{O}_{+}\rangle$ in terms of temperature for
different values of $\alpha $ and $b$ (see Fig. \ref{fig2}). These curves
are obtained by the shooting method which we described in the previous
section. As one can see from this figure there is a critical temperature $%
T_{c}$ below which the condensate appears, then rises quickly as the system
is cooled and finally goes to a constant for sufficiently low temperatures.
This behaviour is qualitatively similar to that obtained in BCS theory and
observed in many materials.
\begin{table}[tbp]
\centering
\begin{tabular}{ccccccc}
\hline\hline
$b$ & $a $ & $\lambda_{\mathrm{min}}^{2} $ & $\gamma_{SL}$ & $\gamma_{%
\mathrm{Numerical}}$ &  &  \\[0.05ex] \hline
0 & 0.721772 & 18.2331 & 7.70525 & 12.4879 &  &  \\
0.01 & 0.747560 & 25.0682 & 9.05565 & 17.4138 &  &  \\
0.02 & 0.777479 & 36.1392 & 14.4103 & 24.5663 &  &  \\[0.5ex] \hline
\end{tabular}%
\caption{The analytical and numerical results for the condensation operator
for $\protect\alpha =0.0001.$}
\end{table}
\begin{table}[tbp]
\centering
\begin{tabular}{ccccccc}
\hline\hline
$b$ & $a $ & $\lambda_{\mathrm{min}}^{2} $ & $\gamma_{SL}$ & $\gamma_{%
\mathrm{Numerical}}$ &  &  \\[0.05ex] \hline
0 & 0.720561 & 18.5392 & 7.72419 & 11.2411 &  &  \\
0.01 & 0.747087 & 25.6427 & 9.44846 & 15.9088 &  &  \\
0.02 & 0.777900 & 37.2577 & 15.6715 & 23.5892 &  &  \\[0.5ex] \hline
\end{tabular}%
\caption{The analytical and numerical results for the condensation operator
for $\protect\alpha =0.01.$}
\end{table}
\begin{table}[tbp]
\centering
\begin{tabular}{ccccccc}
\hline\hline
$b$ & $a $ & $\lambda_{\mathrm{min}}^{2} $ & $\gamma_{SL}$ & $\gamma_{%
\mathrm{Numerical}}$ &  &  \\[0.05ex] \hline
0 & 0.709061 & 21.5679 & 7.90303 & 4.05071 &  &  \\
0.01 & 0.743348 & 31.6982 & 9.81189 & 10.0376 &  &  \\
0.02 & 0.783517 & 49.9342 & 37.6568 & 17.7555 &  &  \\[0.5ex] \hline
\end{tabular}%
\caption{The analytical and numerical results for the condensation operator
for $\protect\alpha =0.1.$}
\end{table}
Now we are going to study the condensation operator $\langle\mathcal{O}%
_{+}\rangle$ in the close neighborhood of the superconductor critical
temperature to compute the critical exponents and the condensation value $%
\gamma$ of the Gauss-Bonnet HSC with quadratic nonlinear electromagnetic.
For this purpose, we first take the logarithmic of Eq. (\ref{gamma}). We
arrive at
\begin{equation}
\mathcal{\log }\left( \frac{\langle \mathcal{O_{+}}\rangle }{T_{c}^{3}}%
\right) =\mathcal{\log }\left( \pi ^{3}\gamma \right) +\frac{1}{2}\mathcal{%
\log }\left( 1-\frac{T}{T_{c}}\right) .
\end{equation}%
\begin{figure*}[t]
\begin{center}
\begin{minipage}[b]{0.325\textwidth}\begin{center}
       \subfigure[~$\alpha=0.0001$]{
                \label{fig3a}\includegraphics[width=\textwidth]{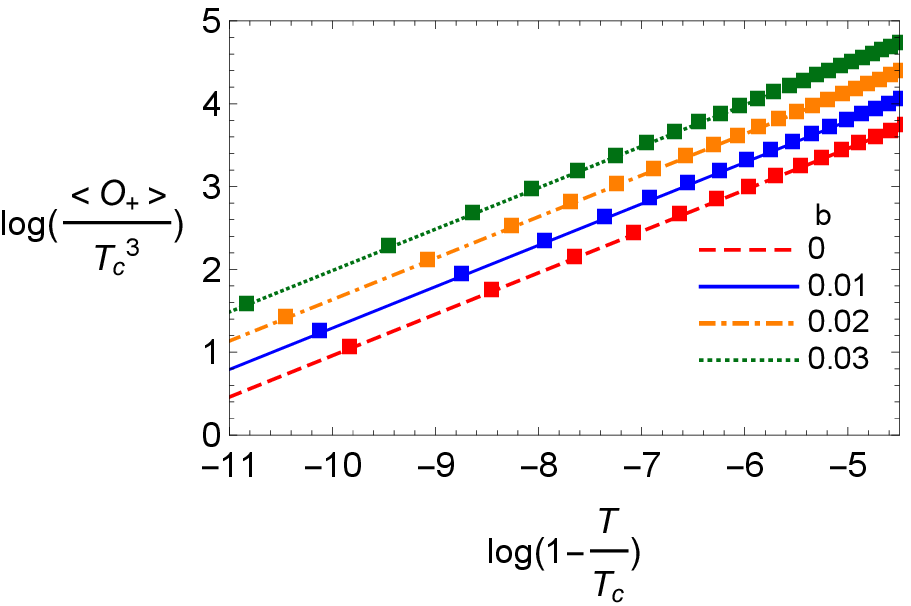}\qquad}
    \end{center}\end{minipage} \hskip+0cm
\begin{minipage}[b]{0.325\textwidth}\begin{center}
        \subfigure[~$\alpha=0.01$]{
                 \label{fig3b}\includegraphics[width=\textwidth]{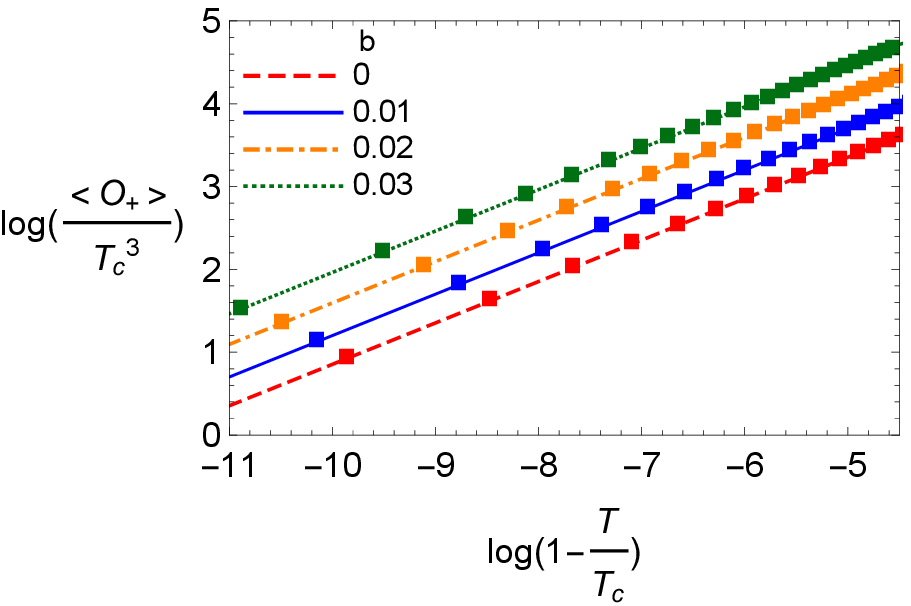}\qquad}
    \end{center}\end{minipage} \hskip0cm
\begin{minipage}[b]{0.325\textwidth}\begin{center}
         \subfigure[~$\alpha=0.1$]{
                  \label{fig3c}\includegraphics[width=\textwidth]{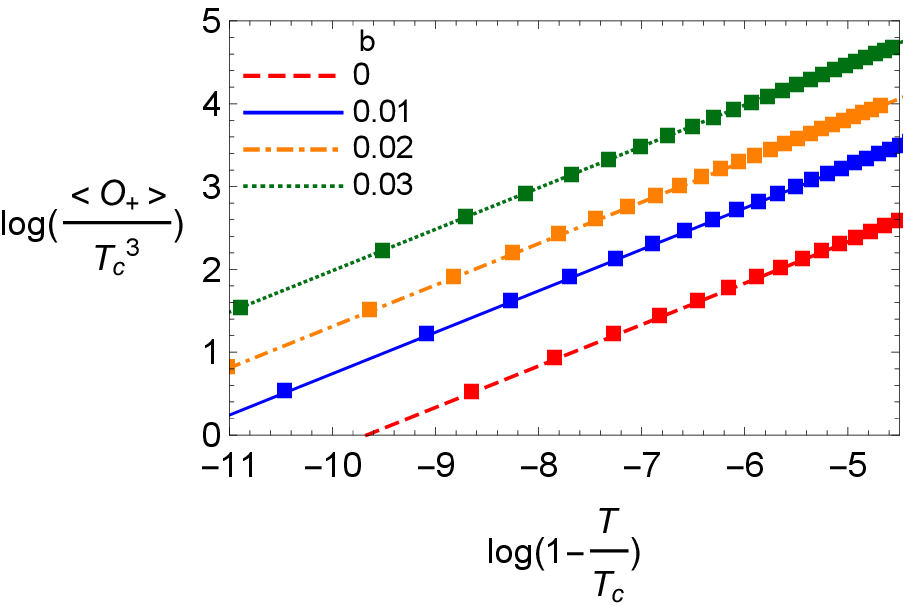}\qquad}
    \end{center}\end{minipage} \hskip0cm
\end{center}
\caption{These figures show the behavior of $\log{\left(<\mathcal{O}
_{+}>/T_{c}^3\right)}$ versus $\log\left(1-T/Tc\right)$ near the critical
temperature $Tc$ for different values of $b$ and $\protect\alpha $. The
numerical results are highlighted in the filled squares.}
\label{fig3}
\end{figure*}
We have plotted the behaviour of the above function in Fig. (\ref{fig3}).
From this figure, we observe that the numerical results are fitted to the
above analytic form in the vicinity of the critical temperature. We
summarize our results in Fig. (\ref{fig3}) and also tables $4-6$ for
different values of $b$ and $\alpha $. We see that for a fixed value of $%
\alpha $, the condensation operator $\gamma $ increases with increasing $b$,
while for a fixed value of $b$, it decreases with increasing $\alpha $.

\section{Holographic Conductivity \label{Cond}}

In this section, we study the energy gap in the holographic superconductor
phase which is constructed on the boundary of the background spacetime. In
particular, we investigate the influence of the Gauss-Bonnet and nonlinear
parameters on the superconducting gap. In order to do this, we must compute
the electrical conductivity of holographic superconductor by turning on a
small perturbation $\delta A_{x}=A_{x}(r)\exp (-i\omega t)$ to the gauge
field in the bulk where $\omega $ is the frequency. At linearized order in
perturbation ($\delta A_{x}$), the equation of motion for the gauge field $%
A_{x}(r)$, which obeys Eq. (\ref{FE3}), is
\begin{equation}
\partial _{r}^{2}A_{x}+\left[ \frac{1}{r}\left( 1-6b\partial _{r}\phi
^{2}\right) +\frac{\partial _{r}f}{f}+\frac{4b\phi \partial _{r}\phi \psi
^{2}}{f}\right] \partial _{r}A_{x}-\left[ \frac{2\psi ^{2}}{f}\left(
1-b\partial _{r}\phi ^{2}\right) -\frac{\omega ^{2}}{f^{2}}\right] A_{x}=0.
\label{eqAx}
\end{equation}%
In the absence of the nonlinear correction ($b=0$), this differential
equation reduces to the Maxwell case as presented in \cite{Hartnol,Hartnol2}%
. To determine the conductivity, we need the asymptotic ($r\rightarrow
\infty $) form of the second order differential equation (\ref{eqAx}), which
may be obtained as%
\begin{equation}
\partial _{r}^{2}A_{x}+\frac{3}{r}\partial _{r}A_{x}+\frac{\omega ^{2}L_{%
\mathrm{eff}}^{4}}{r^{4}}A_{x}+\ldots =0,
\end{equation}%
which admits the following solution near the boundary
\begin{equation}
A_{x}=A_{x}^{(0)}+\frac{A_{x}^{(1)}}{r^{2}}+\frac{\omega ^{2}L_{\mathrm{eff}%
}^{4}\mathrm{ln}(Kr)}{2r^{2}}A_{x}^{(0)}+\cdots ,  \label{asympAx}
\end{equation}%
where $A_{x}^{(0)}$, $A_{x}^{(1)}$ are two constant and $K$ is also a
constant parameter with length dimension which is considered for a
dimensionless logarithmic argument.

According to AdS/CFT correspondence, the two point correlation function of
the current operators in a system is given by its on shell action where the
action is evaluated on the equations of motion. Here, the on shell action is
\begin{equation}
S_{o.s.}\equiv \int_{r_{+}}^{r_{\infty }}dr\int d^{4}x\sqrt{-g}\mathcal{L},
\end{equation}%
which, in the quadratic approximation for the gauge field perturbation
becomes%
\begin{equation}
S_{o.s.}=\int d^{4}x\int_{r_{+}}^{r_{\infty }}dr\Bigg\{-\frac{1}{2}r\left[
\left( 2\psi (r)^{2}-\frac{\omega ^{2}}{f(r)}-\frac{b\omega ^{2}\partial
_{r}\phi ^{2}}{f(r)}\right) A_{x}(r)^{2}+f(r)\left( 1+b\partial _{r}\phi
^{2}\right) \partial _{r}A_{x}(r)^{2}\right] \Bigg\}.
\end{equation}%
After performing an integration by parts and using Eq. (\ref{eqAx}), we get%
\begin{equation}
S_{o.s.}=\int d^{4}x\left[ -\frac{1}{2}rf(r)\left( 1+b\partial _{r}\phi
^{2}\right) \partial _{r}A_{x}(r)A_{x}(r)\right] \Bigg|_{r=r_{\infty }}.
\end{equation}%
\textbf{Substituting Eqs. (\ref{f}), (\ref{phi}) and
(\ref{asympAx})
in the above expression, one arrives at}%
\begin{eqnarray}
&&S_{o.s.}=\int d^{4}x\left[ \frac{A_{x}^{(0)}A_{x}^{(1)}}{L_{\mathrm{eff}%
}^{2}}-\frac{\omega ^{2}L_{\mathrm{eff}}^{2}A_{x}^{(0)2}}{4}+\frac{1}{2}%
\omega ^{2}L_{\mathrm{eff}}^{2}\mathrm{ln}(Kr)A_{x}^{(0)2}+\frac{A_{x}^{(1)2}%
}{L_{\mathrm{eff}}^{2}r^{2}}\right.   \notag \\
&&-\frac{\omega ^{2}L_{\mathrm{eff}}^{2}A_{x}^{(0)}A_{x}^{(1)}}{4r^{2}}-%
\frac{\omega ^{2}L_{\mathrm{eff}}^{6}\mathrm{ln}(Kr)A_{x}^{(0)2}}{8r^{2}}+%
\frac{\omega ^{2}L_{\mathrm{eff}}^{2}\mathrm{ln}(Kr)A_{x}^{(0)}A_{x}^{(1)}}{%
r^{2}}  \notag \\
&&\left. +\frac{\omega ^{4}L_{\mathrm{eff}}^{6}\mathrm{ln}(Kr)A_{x}^{(0)2}}{%
4r^{2}}+\mathcal{O}\left(\frac{1}{r^{3}}\right)\right]
\Bigg|_{r=r_{\infty }},
\end{eqnarray}%
\textbf{thus we can obtain }$S_{o.s.}$\textbf{\ as follows }%
\begin{equation}
S_{o.s.}=\int d^{4}x\left[ \frac{A_{x}^{(0)}A_{x}^{(1)}}{L_{\mathrm{eff}}^{2}%
}-\frac{\omega ^{2}L_{\mathrm{eff}}^{2}A_{x}^{(0)2}}{4}+\frac{1}{2}\omega
^{2}L_{\mathrm{eff}}^{2}\mathrm{ln}(Kr)A_{x}^{(0)2}\right] ,
\end{equation}%
\textbf{in which logarithmic divergences appears. In order to cancel out
this divergency, we obtain the boundary counterterm as described in Appendix by using Skenderis's method of holographic renormalization \cite{Sken}. }%
Therefore, the finite on shell action may be written as%
\begin{equation}
S=S_{o.s.}+S_{c.t.},
\end{equation}%
\begin{figure*}[t]
\begin{center}
\begin{minipage}[b]{0.325\textwidth}\begin{center}
       \subfigure[~$\alpha=0.0001,b=0$]{
                \label{fig4a}\includegraphics[width=\textwidth]{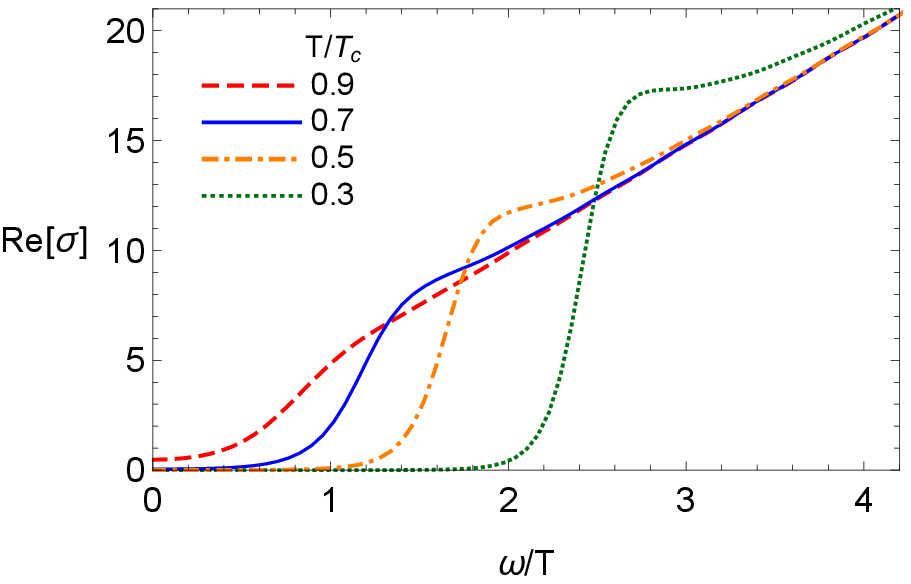}\qquad}
    \end{center}\end{minipage} \hskip+0cm
\begin{minipage}[b]{0.325\textwidth}\begin{center}
        \subfigure[~$\alpha=0.0001,b=0.01$]{
                 \label{fig4b}\includegraphics[width=\textwidth]{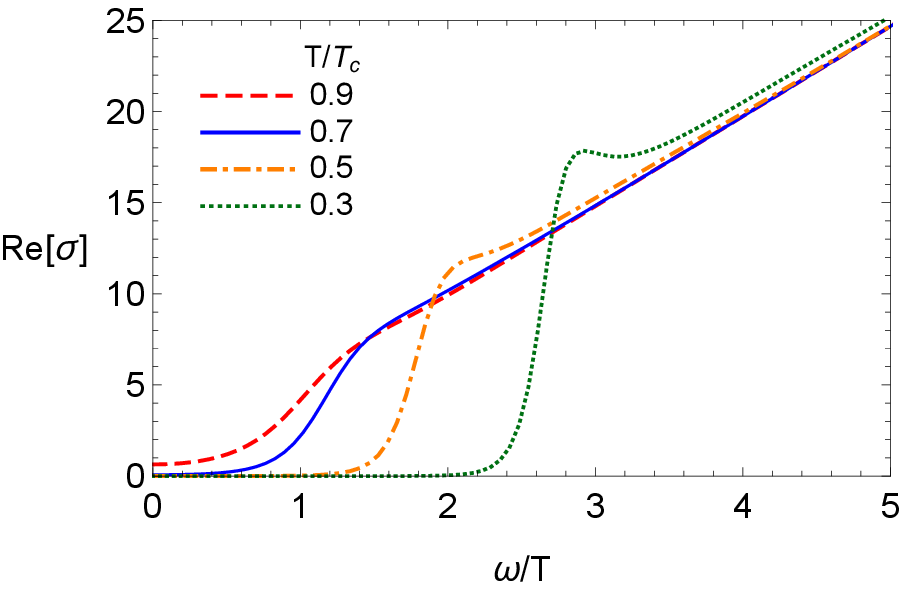}\qquad}
    \end{center}\end{minipage} \hskip0cm
\begin{minipage}[b]{0.325\textwidth}\begin{center}
         \subfigure[~$\alpha=0.0001,b=0.02$]{
                  \label{fig4c}\includegraphics[width=\textwidth]{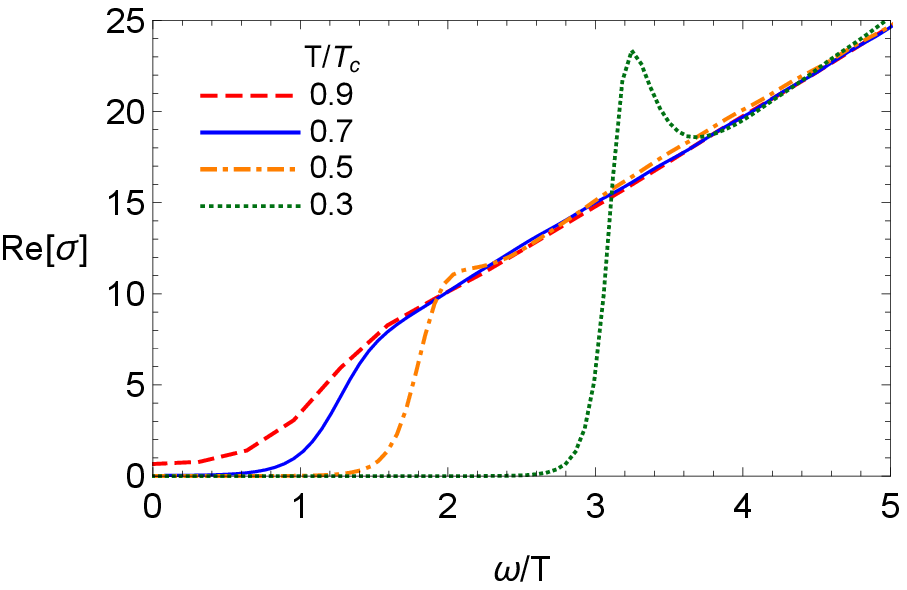}\qquad}
    \end{center}\end{minipage} \hskip0cm
\begin{minipage}[b]{0.325\textwidth}\begin{center}
             \subfigure[~$\alpha=0.01,b=0$]{
                      \label{fig4d}\includegraphics[width=\textwidth]{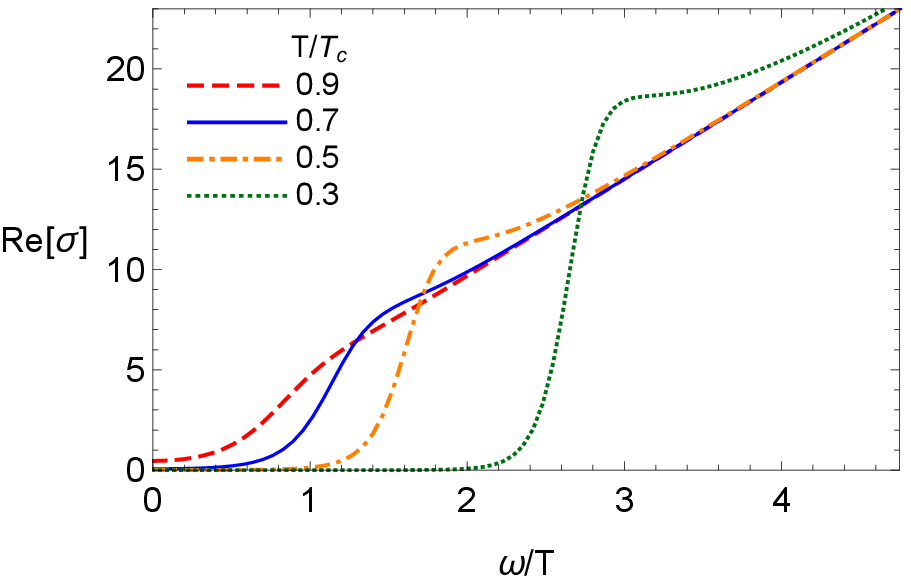}\qquad}
        \end{center}\end{minipage} \hskip0cm
\begin{minipage}[b]{0.325\textwidth}\begin{center}
                     \subfigure[~$\alpha=0.01,b=0.01$]{
                                           \label{fig4e}\includegraphics[width=\textwidth]{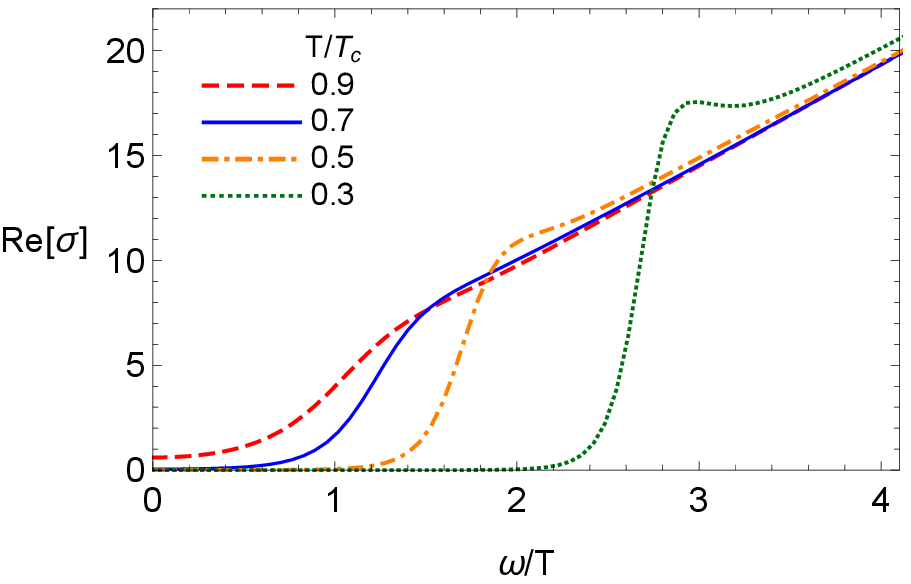}\qquad}
        \end{center}\end{minipage} \hskip0cm
\begin{minipage}[b]{0.325\textwidth}\begin{center}
       \subfigure[~$\alpha=0.1,b=0$]{
                \label{fig4f}\includegraphics[width=\textwidth]{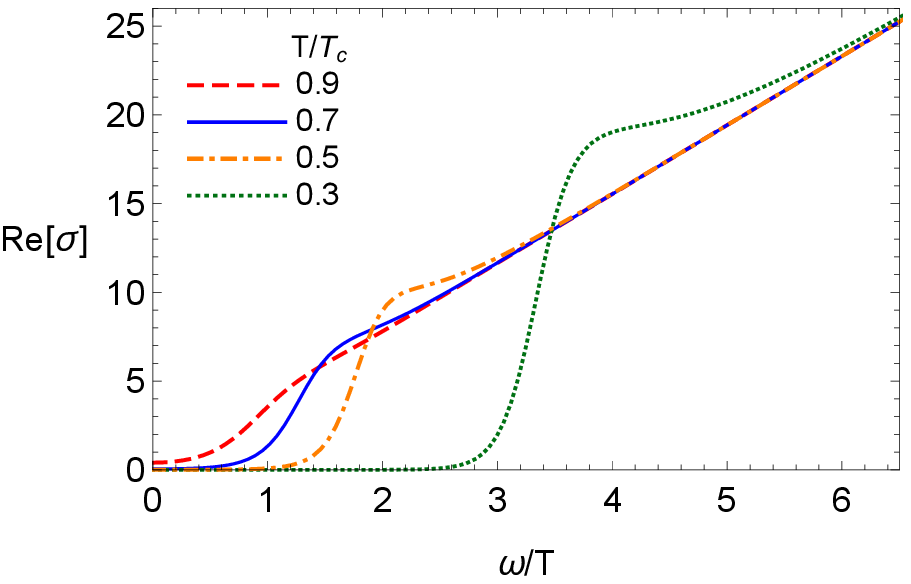}\qquad}
    \end{center}\end{minipage} \hskip+0cm
\begin{minipage}[b]{0.325\textwidth}\begin{center}
       \subfigure[~$\alpha=0.1,b=0.01$]{
                \label{fig4g}\includegraphics[width=\textwidth]{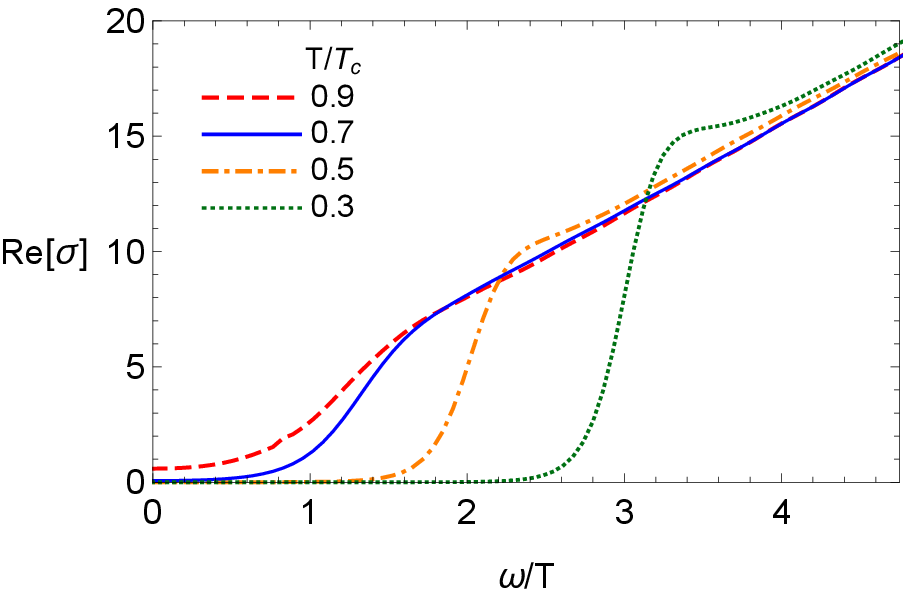}\qquad}
    \end{center}\end{minipage} \hskip+0cm
\end{center}
\caption{The real part of the conductivity as a function of frequency for
different values of $b$ and $\protect\alpha $. Each figure is plotted for
different temperature $T/Tc$.}
\label{fig4}
\end{figure*}
\begin{figure*}[t]
\begin{center}
\begin{minipage}[b]{0.325\textwidth}\begin{center}
       \subfigure[~$%
               \alpha=0.0001,b=0$]{
                \label{fig5a}\includegraphics[width=\textwidth]{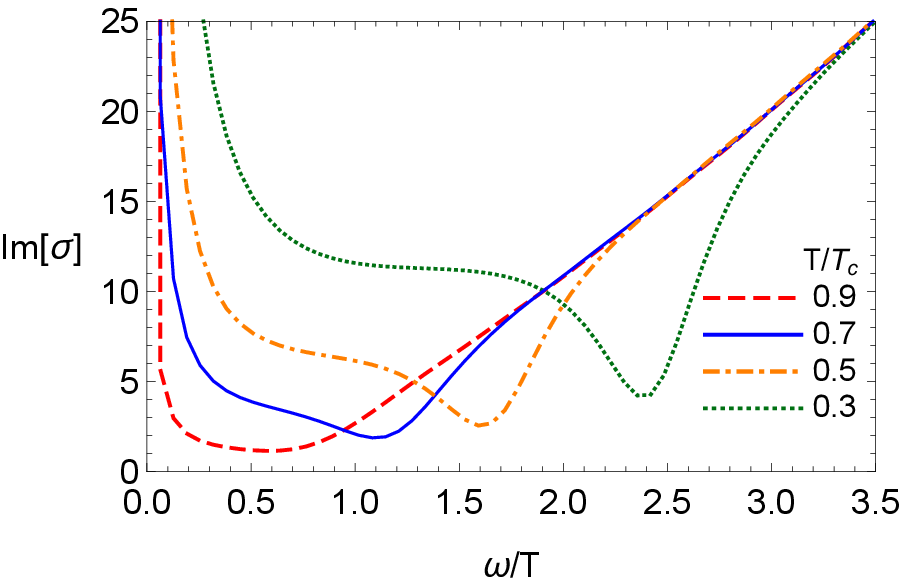}\qquad}
    \end{center}\end{minipage} \hskip+0cm
\begin{minipage}[b]{0.325\textwidth}\begin{center}
        \subfigure[~$\alpha=0.0001,b=0.01$]{
                 \label{fig5b}\includegraphics[width=\textwidth]{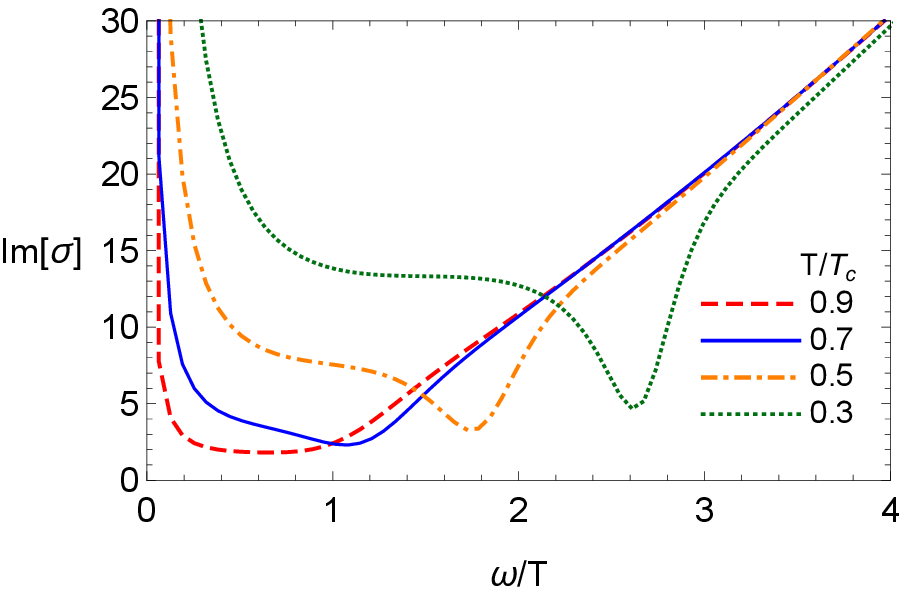}\qquad}
    \end{center}\end{minipage} \hskip0cm
\begin{minipage}[b]{0.325\textwidth}\begin{center}
         \subfigure[~$\alpha=0.0001,b=0.02$]{
                  \label{fig5c}\includegraphics[width=\textwidth]{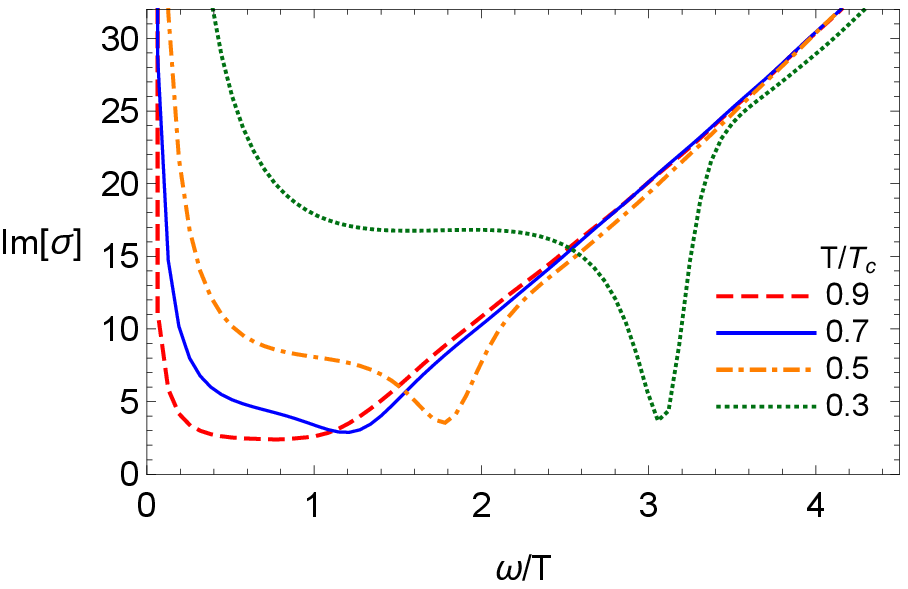}\qquad}
    \end{center}\end{minipage} \hskip0cm
\begin{minipage}[b]{0.325\textwidth}\begin{center}
             \subfigure[~$%
                                  \alpha=0.01,b=0$]{
                      \label{fig5d}\includegraphics[width=\textwidth]{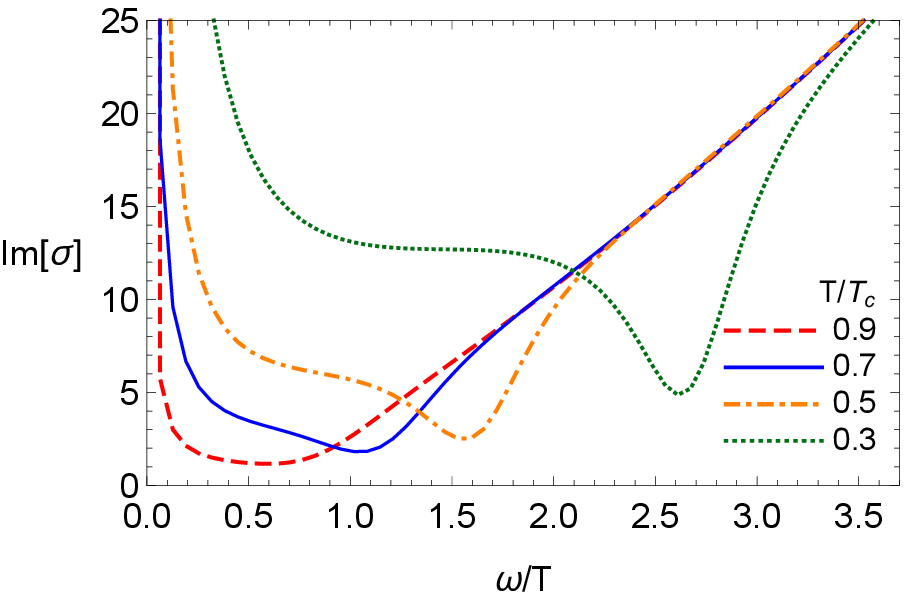}\qquad}
        \end{center}\end{minipage} \hskip0cm
\begin{minipage}[b]{0.325\textwidth}\begin{center}
                     \subfigure[~$\alpha=0.01,b=0.01$]{
                                           \label{fig5e}\includegraphics[width=\textwidth]{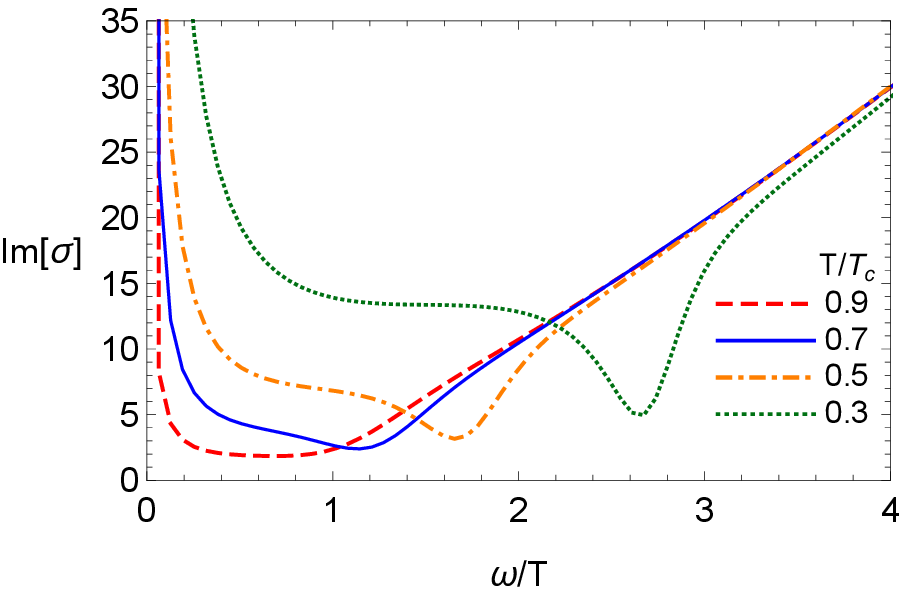}\qquad}
        \end{center}\end{minipage} \hskip0cm
\begin{minipage}[b]{0.325\textwidth}\begin{center}
       \subfigure[~$\alpha=0.1,b=0$]{
                \label{fig5f}\includegraphics[width=\textwidth]{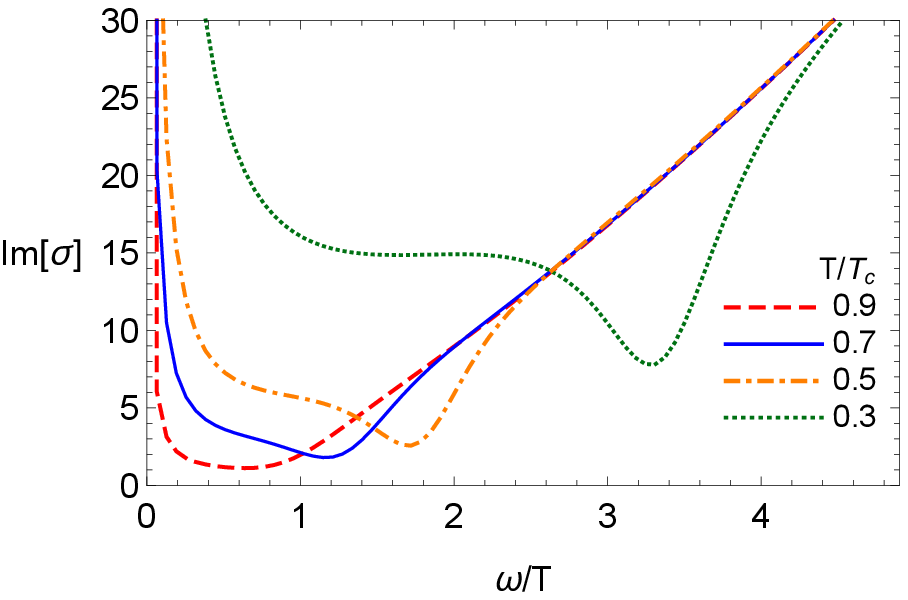}\qquad}
    \end{center}\end{minipage} \hskip+0cm
\begin{minipage}[b]{0.325\textwidth}\begin{center}
       \subfigure[~$\alpha=0.1,b=0.01$]{
                \label{fig5g}\includegraphics[width=\textwidth]{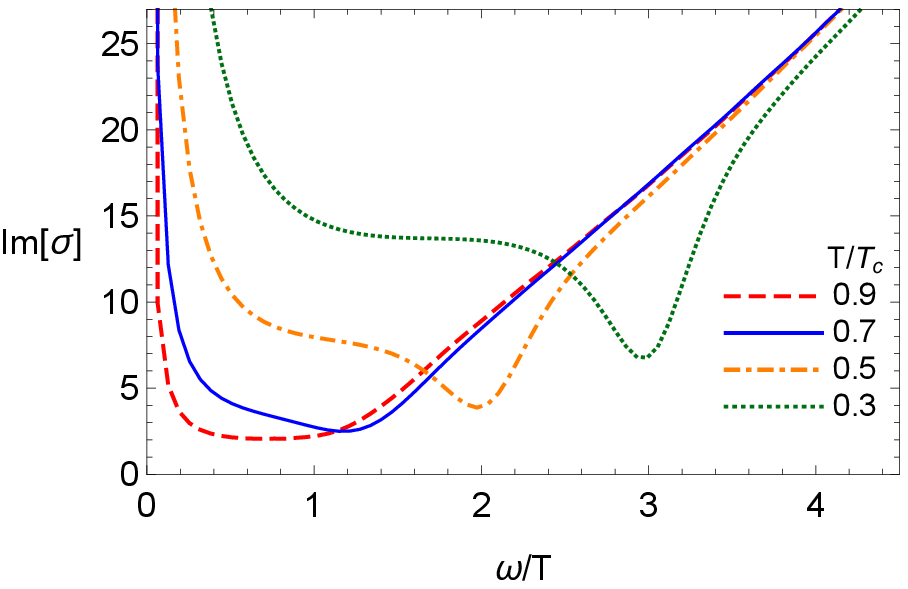}\qquad}
    \end{center}\end{minipage} \hskip+0cm
\end{center}
\caption{The imaginary part of the conductivity as a function of frequency
for different values of $b$ and $\protect\alpha $. Each figure is plotted
for different temperature $T/Tc$.}
\label{fig5}
\end{figure*}
\textbf{where  the gauge invariant
counterterm $S_{c.t.}$ is given by Eq. ({\ref{SCt}})}. Now, we can obtain the current operator in the boundary field theory \cite%
{Hartnol,Hartnol2} as%
\begin{equation}
\left\langle J_{x}\right\rangle =\frac{\delta S}{\delta A_{x}^{(0)}}=\frac{2%
}{L_{\mathrm{eff}}^{2}}A_{x}^{(1)}-\frac{\omega ^{2}L_{\mathrm{eff}}^{2}}{2}%
A_{x}^{(0)}\text{ }.  \label{current}
\end{equation}%
According to Ohm's law, the electrical conductivity can be expressed as%
\begin{equation}
\sigma \left( \omega \right) =\frac{\left\langle J_{x}\right\rangle }{E_{x}},
\end{equation}%
where $E_{x}=-\partial _{t}\delta A_{x}$. Hence, using the current (\ref%
{current}), the holographic conductivity is given by%
\begin{equation}
\sigma =-\frac{2iA_{x}^{(1)}}{\omega L_{\mathrm{eff}}^{2}A_{x}^{(0)}}+\frac{%
i\omega L_{\mathrm{eff}}^{2}}{2}.
\end{equation}%
Consequently, the holography conductivity is calculated by solving
numerically a differential equation (\ref{eqAx}) such that the infalling
boundary condition is imposed at the event horizon%
\begin{equation}
A_{x}(r)=\mathrm{exp}\left( -\frac{i\omega }{4\pi T}\right) S(r),
\end{equation}%
in which $T$ is the Hawking temperature and%
\begin{equation}
S(r)=1+a_{1}(r-r_{+})+a_{2}(r-r_{+})^{2}+\ldots ,
\end{equation}%
where $a_{1},a_{2},\ldots $ are calculated by Taylor series expansion of
equation (\ref{eqAx}) around the horizon.
\begin{figure*}[t]
\begin{center}
\begin{minipage}[b]{0.325\textwidth}\begin{center}
       \subfigure[~$\alpha=0.0001$]{
                \label{fig6a}\includegraphics[width=\textwidth]{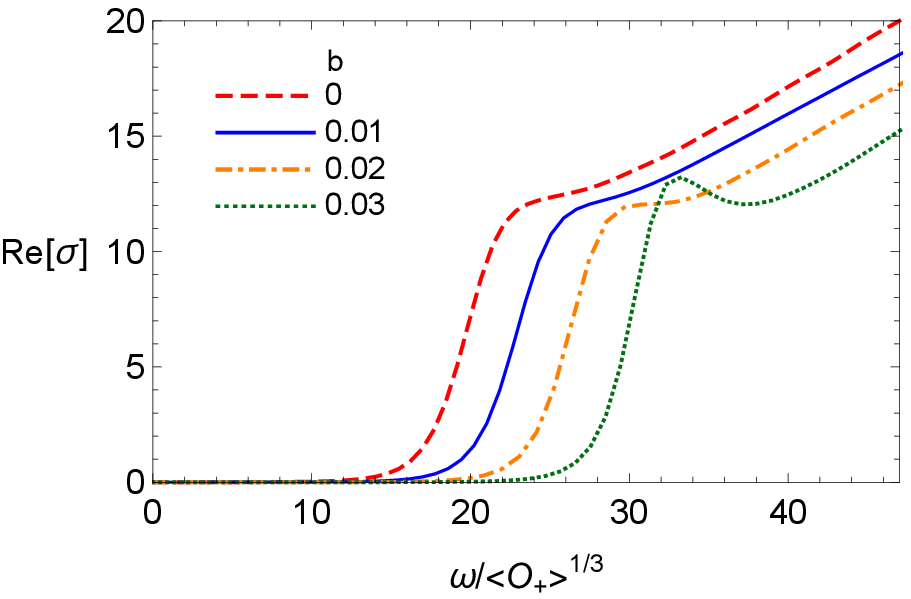}\qquad}
    \end{center}\end{minipage} \hskip+0cm
\begin{minipage}[b]{0.325\textwidth}\begin{center}
        \subfigure[~$%
        \alpha=0.01$]{
                 \label{fig6b}\includegraphics[width=\textwidth]{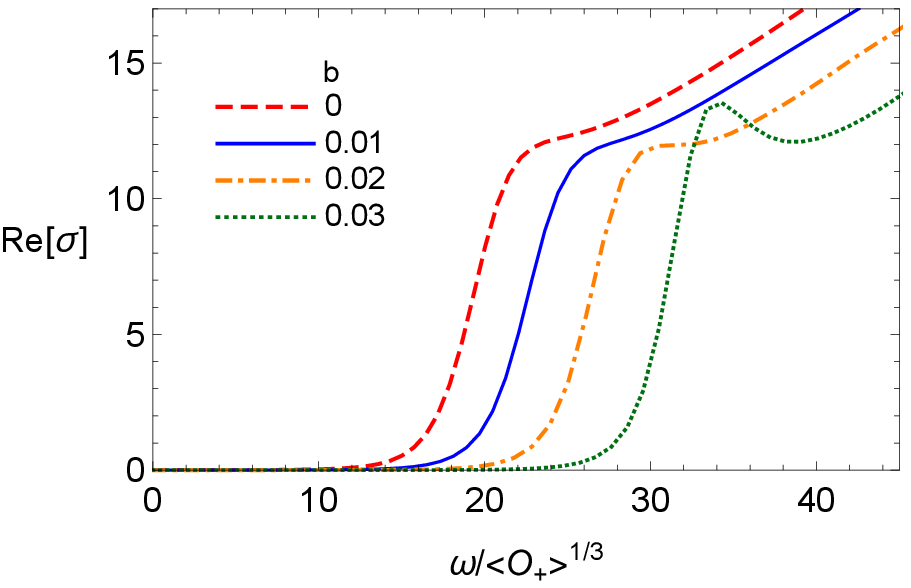}\qquad}
    \end{center}\end{minipage} \hskip0cm
\begin{minipage}[b]{0.325\textwidth}\begin{center}
         \subfigure[~$\alpha=0.1$]{
                  \label{fig6c}\includegraphics[width=\textwidth]{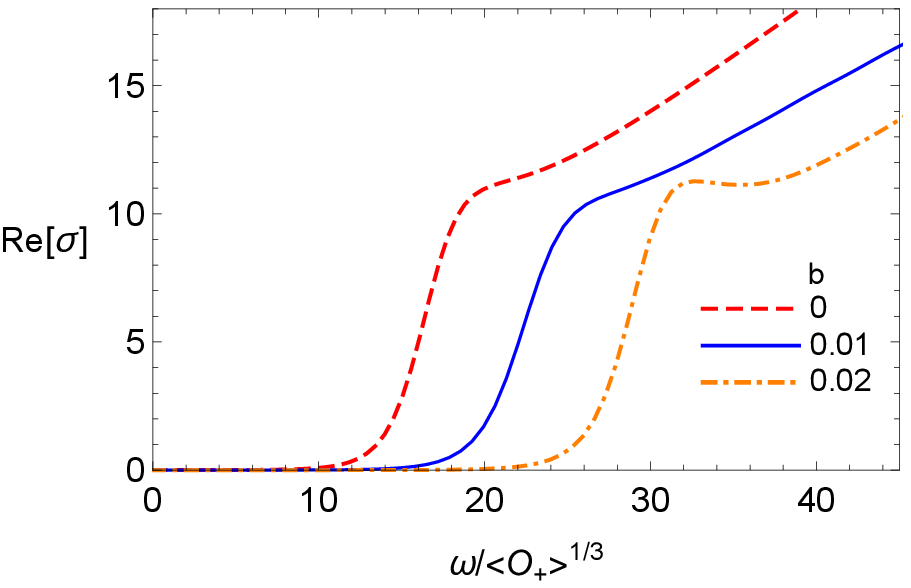}\qquad}
    \end{center}\end{minipage} \hskip0cm
\begin{minipage}[b]{0.325\textwidth}\begin{center}
         \subfigure[~$b=0$]{
                  \label{fig6d}\includegraphics[width=\textwidth]{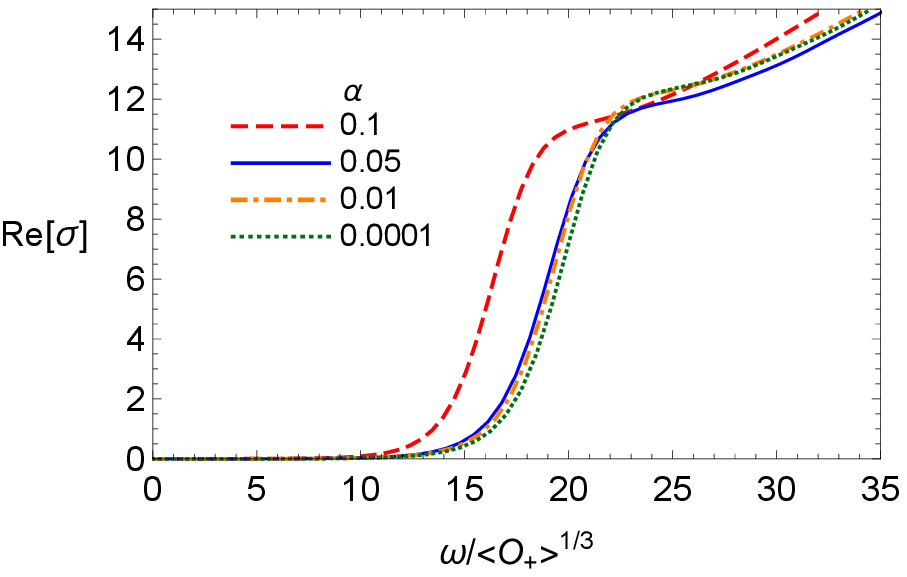}\qquad}
    \end{center}\end{minipage} \hskip0cm
\begin{minipage}[b]{0.325\textwidth}\begin{center}
         \subfigure[~$b=0.005$]{
                  \label{fig6e}\includegraphics[width=\textwidth]{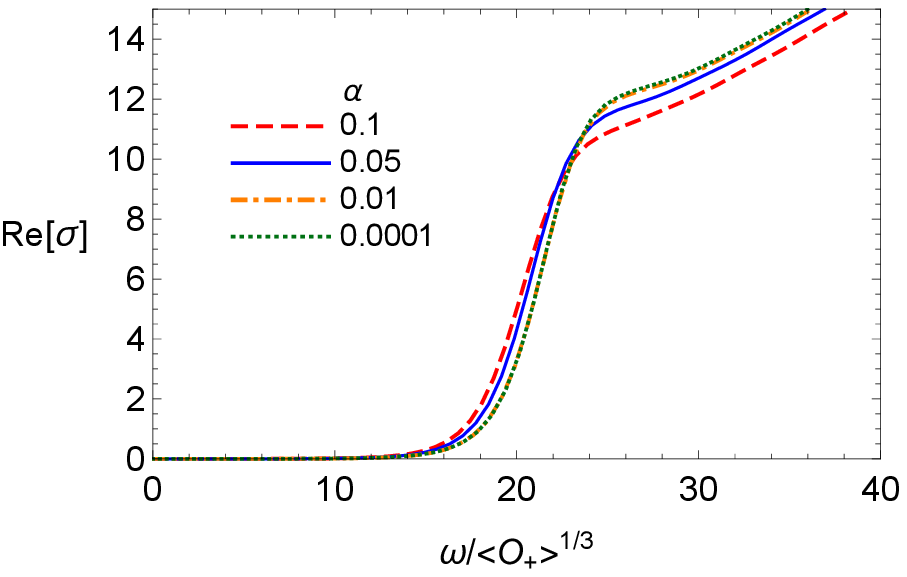}\qquad}
    \end{center}\end{minipage} \hskip0cm
\begin{minipage}[b]{0.325\textwidth}\begin{center}
             \subfigure[~$b=0.01$]{
                      \label{fig6f}\includegraphics[width=\textwidth]{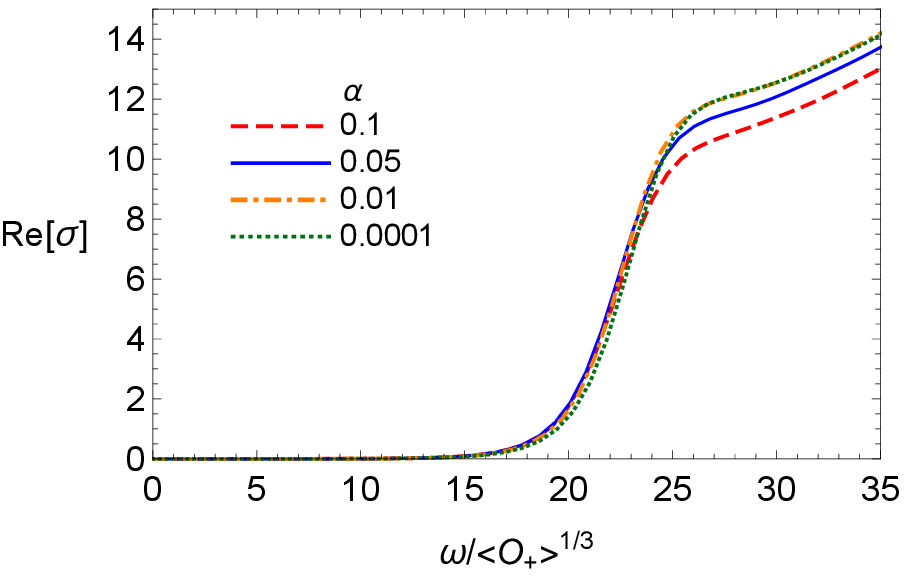}\qquad}
        \end{center}\end{minipage} \hskip0cm
\begin{minipage}[b]{0.325\textwidth}\begin{center}
                     \subfigure[~$b=0.02$]{
                              \label{fig6g}\includegraphics[width=\textwidth]{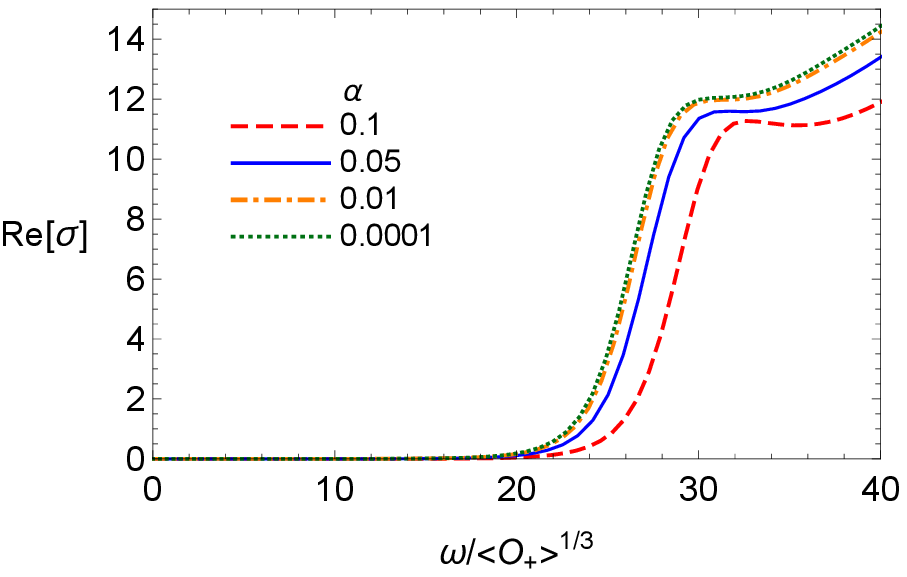}\qquad}
                \end{center}\end{minipage} \hskip0cm
\end{center}
\caption{The real part of the conductivity as a function of $\protect\omega %
/<\mathcal{O_{+}}>^{1/3}$, at low temperatures, around $T\approx 0.1Tc$.}
\label{fig6}
\end{figure*}
\begin{figure*}[t]
\begin{center}
\begin{minipage}[b]{0.325\textwidth}\begin{center}
       \subfigure[~$\alpha=0.0001$]{
                \label{fig7a}\includegraphics[width=\textwidth]{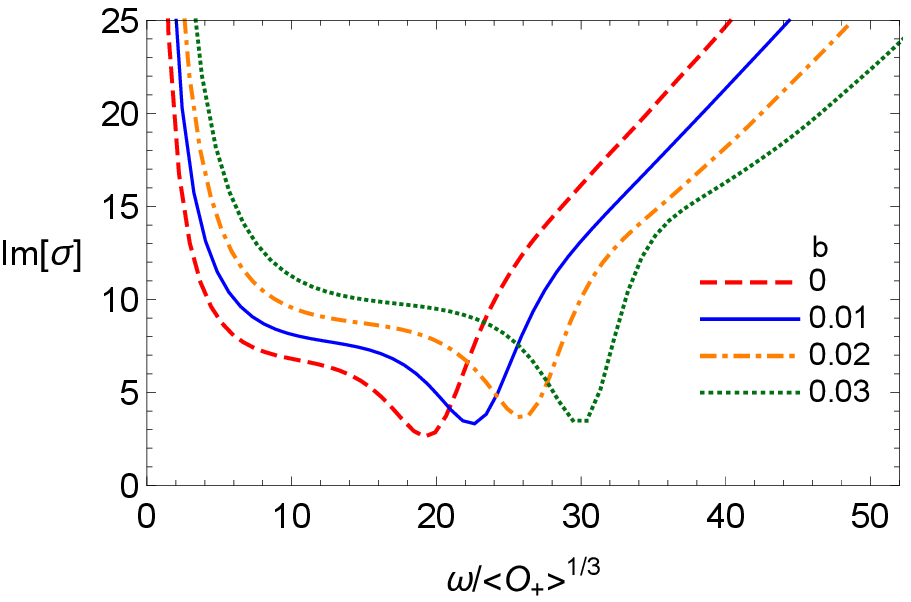}\qquad}
    \end{center}\end{minipage} \hskip+0cm
\begin{minipage}[b]{0.325\textwidth}\begin{center}
        \subfigure[$%
        \alpha=0.01$]{
                 \label{fig7b}\includegraphics[width=\textwidth]{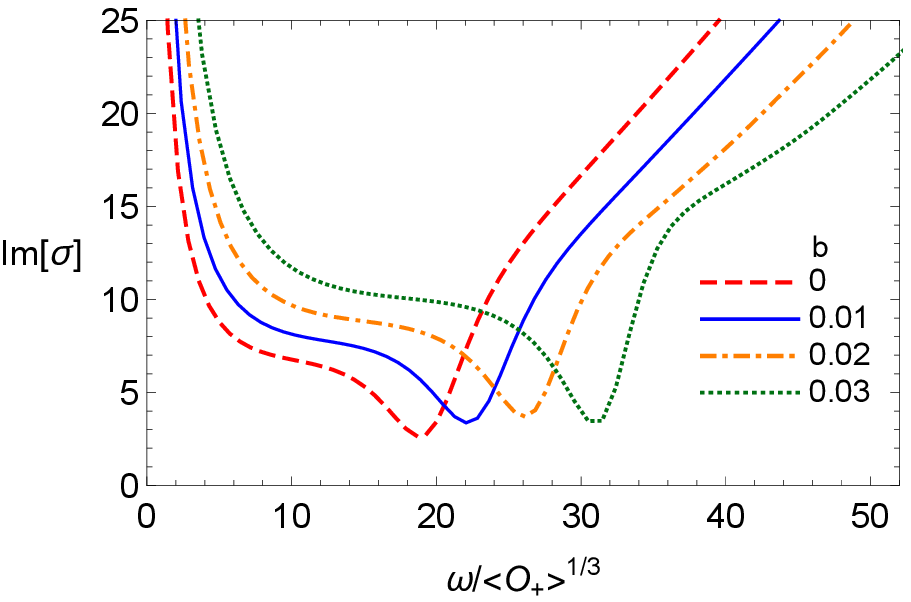}\qquad}
    \end{center}\end{minipage} \hskip0cm
\begin{minipage}[b]{0.325\textwidth}\begin{center}
         \subfigure[~$\alpha=0.1$]{
                  \label{fig7c}\includegraphics[width=\textwidth]{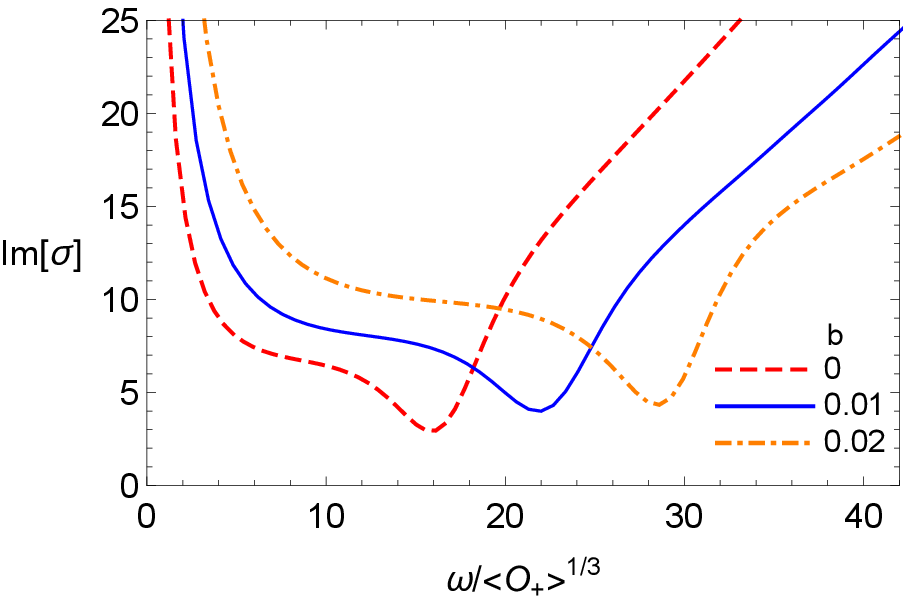}\qquad}
    \end{center}\end{minipage} \hskip0cm
\begin{minipage}[b]{0.325\textwidth}\begin{center}
         \subfigure[~$b=0$]{
                  \label{fig7d}\includegraphics[width=\textwidth]{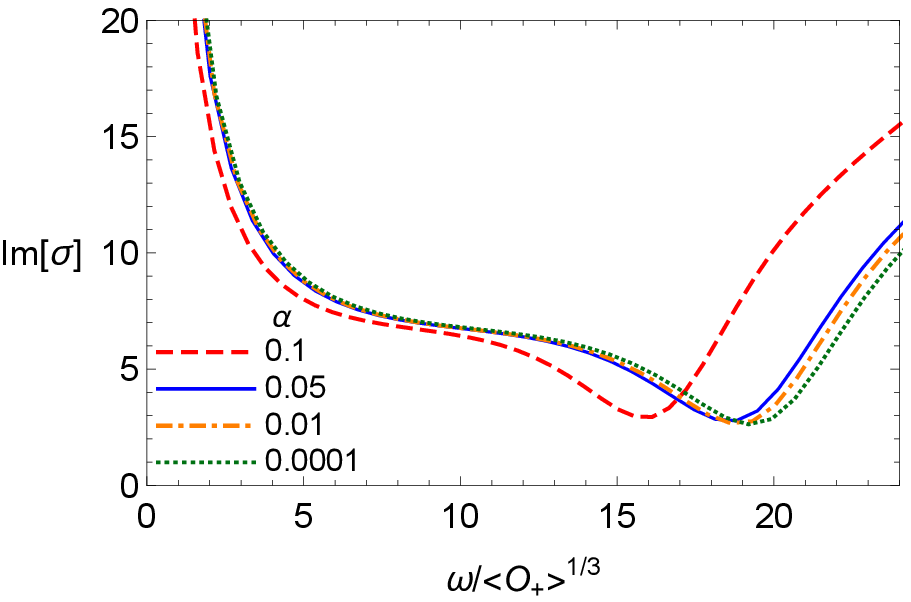}\qquad}
    \end{center}\end{minipage} \hskip0cm
\begin{minipage}[b]{0.325\textwidth}\begin{center}
         \subfigure[~$b=0.005$]{
                  \label{fig7e}\includegraphics[width=\textwidth]{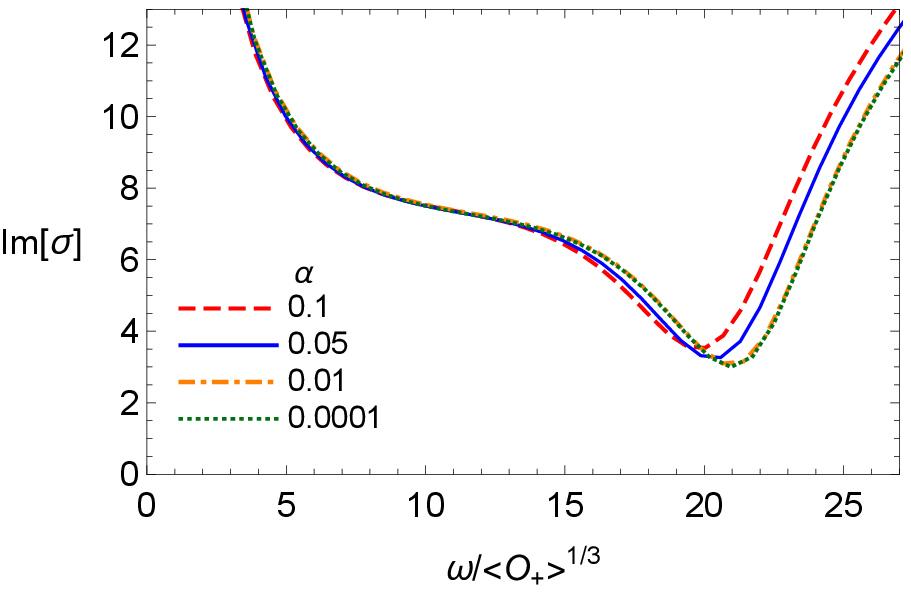}\qquad}
    \end{center}\end{minipage} \hskip0cm
\begin{minipage}[b]{0.325\textwidth}\begin{center}
             \subfigure[~$b=0.01$]{
                      \label{fig7f}\includegraphics[width=\textwidth]{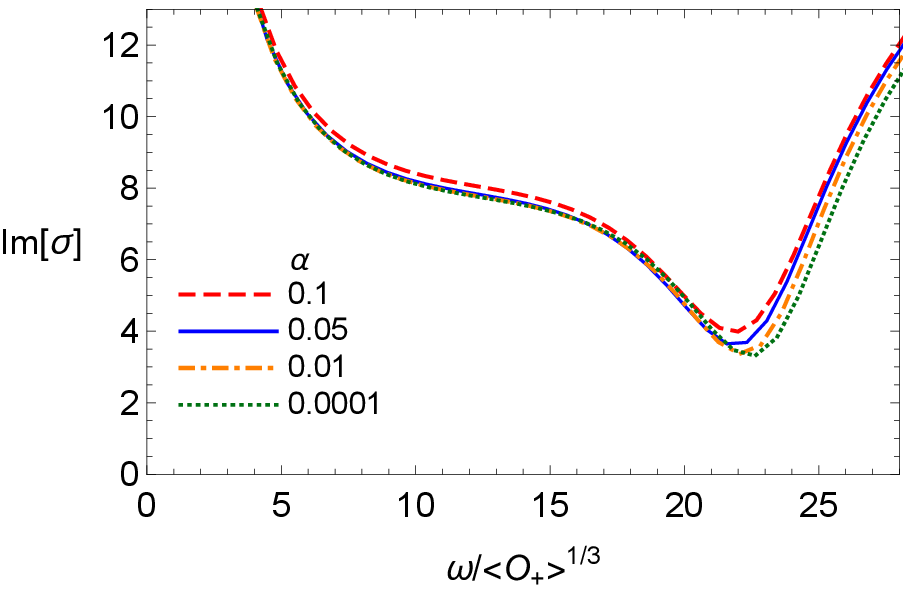}\qquad}
        \end{center}\end{minipage} \hskip0cm
\begin{minipage}[b]{0.325\textwidth}\begin{center}
                 \subfigure[~$b=0.02$]{
                          \label{fig7g}\includegraphics[width=\textwidth]{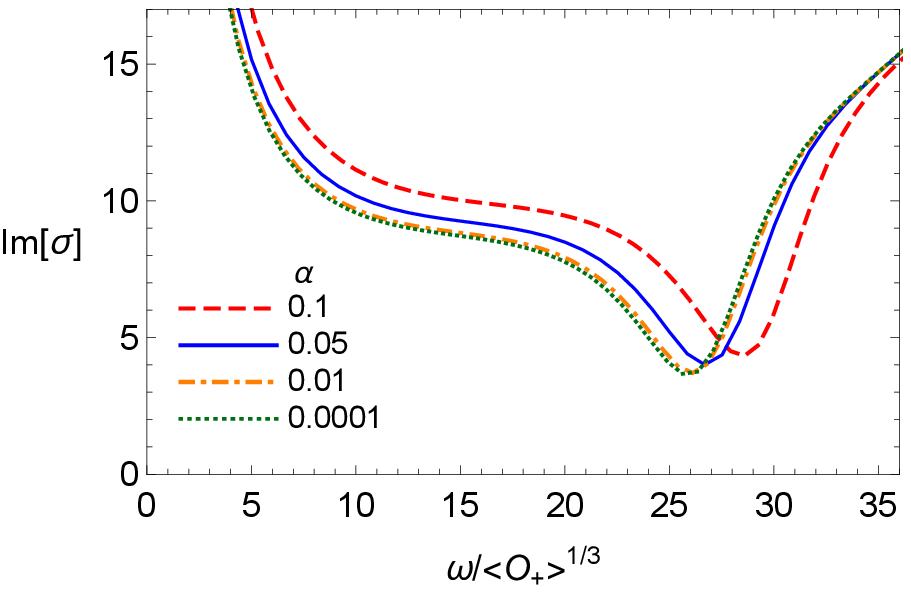}\qquad}
            \end{center}\end{minipage} \hskip0cm
\end{center}
\caption{The imaginary part of the conductivity as a function of $\protect%
\omega /<\mathcal{O_{+}}>^{1/3}$, at low temperatures, around $T\approx 0.1Tc
$. }
\label{fig7}
\end{figure*}

The numerical results for holographic conductivity associated with the
condensation operator $<\mathcal{O_{+}}>$ are plotted in Figs. \ref{fig4}-%
\ref{fig7}. The real and imaginary parts of electrical conductivity versus
frequency are illustrated at different temperature below $Tc$ in Figs.\ref%
{fig4} and \ref{fig5}, respectively. As one can see from Fig. \ref{fig4},
the superconducting gap is opened below the critical temperature which
became deeper with decreasing the temperature. Besides, for various values
of $\alpha $ and $b$, the real part of conductivity is proportional to the
frequency per temperature, $\omega /T$, as frequency is enough large.
According to the Fig. \ref{fig5} the divergence in imaginary part, at $%
\omega =0$, points out a delta function in the real part, $Re[\sigma ]$, at $%
\omega =0$ which is not played in the Fig. \ref{fig4}. As one can see from
Fig. \ref{fig5}, the holographic conductivity of HSC has a minimum in the
imaginary part. Thus with decreasing the temperature the minimum in the
imaginary part shifts toward greater frequency for various values of the
Gauss-Bonnet and nonlinear parameters.

To study formation of the superconducting gap with changing $\alpha $ and $b$
at low temperature, e.g., $T\approx 0.1Tc$, the real and imaginary parts of
holographic conductivity as a function of $\omega /{<\mathcal{O_{+}}>}^{1/3}$
are plotted in Figs. \ref{fig6} and \ref{fig7}, respectively. For a fixed
value of Gauss-Bonnet coefficient $\alpha $, the energy gap ($\omega /{<%
\mathcal{O_{+}}>}^{1/3}$) enlarges with increasing the nonlinear parameter $%
b $. It is evident from Figs. \ref{fig6} and \ref{fig7}, that the energy gap
of HSC for various $\alpha$ exhibits different behavior based on the
nonlinear correction $b$. In case of the Maxwell field ($b=0$), the
superconducting energy gap decreases as $\alpha $ increases at low
temperature (see Fig. \ref{fig6d}). When we take into account the nonlinear
correction $b$, the energy gap of HSC increases with increasing $\alpha $
(see Figs. \ref{fig6d}-\ref{fig6g}). From Fig. \ref{fig7}, we see that for a
fixed value of $\alpha$, the minimum of imaginary part of conductivity goes
to the larger value of $\omega /<\mathcal{O_{+}}>^{1/3}$ when $b$ increases.
In the absence of correction ($b=0$), it decreases with enhancing the
Gauss-Bonnet coefficient (Fig. \ref{fig6d}). Besides, for a fixed value of $%
b $, the minimum of $\mathrm{Im[\sigma]}$ increases with increasing $\alpha$%
, while for a fixed value of $\alpha$, it increase with increasing $b$.

\section{Conclusions}

\label{Con}

In this paper, we continue the studies on the $s$-wave holographic
superconductors (HSC) by taking into account the higher correction terms
both to the gravity side as well as the gauge field side of the system. We
considered the Gauss-Bonnet HSC when the Maxwell Lagrangian has a nonlinear
correction term and is written in the form $\mathcal{L}=\mathcal{F}+b
\mathcal{F}^2$, where $\mathcal{F}$ is the Maxwell lagrangian. We have
provided several motivations for choosing this kind of Lagrangian for the
gauge field. For example, all well-known nonlinear Lagrangian has a series
expansion which their first two terms are exactly in the above form.

First, we have analytically as well as numerically investigated the relation
between critical temperature of phase transition and charge density which
depends on both the Gauss-Bonnet parameter $\alpha$ and the nonlinear
parameter $b$. For this purpose, we employed the analytical Sturm-Liouville
eigenvalue problem and the numerical shooting method. We find out that for a
fixed value of $\alpha$, with increasing the nonlinear parameter $b$, the
value of $T_{c}/\rho ^{1/3}$ decreases as well. This implies that when $b$
becomes larger the condensation gets harder. Similar behavior can be seen
for the fixed value of $b$ and different values of $\alpha $, namely the
critical temperature decreases and the condensation becomes harder when the
Gauss-Bonnet coupling parameter $\alpha$ gets larger. We confirmed that this
results are in a very good agreement with our numerical results. Then, we
obtained the critical exponent of the Gauss-Bonnet HSC with nonlinear gauge
field. We observed that the critical exponent has the mean field value $1/2$%
, which is independent of the nonlinear parameter $b$ and Gauss-Bonnet
parameter $\alpha$.

Then, we explored, numerically, the holographic conductivity of the system.
For this purpose, we plotted the real and imaginary parts of electrical
conductivity versus $\omega /T$ and $\omega /<\mathcal{O_{+}}>^{1/3}$ for $%
T<Tc$. We observed that the superconducting gap is opened below the critical
temperature which became deeper with decreasing the temperature.
Interstingly enough, we found that for different values of $\alpha $ and $b$%
, and for large frequency, the real part of conductivity is proportional to $%
\omega/T$. We observed that the holographic conductivity of HSC has a
minimum in the imaginary part. Besides, with decreasing the temperature the
minimum in the imaginary part shifts toward greater frequency for various
values of the Gauss-Bonnet parameter $\alpha$ and nonlinear gauge field
parameter $b$. Furthermore, for a fixed value of $b$, the minimum of $%
\mathrm{Im[\sigma]}$ increases with increasing $\alpha$, while for a fixed
value of $\alpha$, it increase with increasing $b$.


\section*{Acknowledgments}

We thank the Research Council of Shiraz University. The work of AS has been
supported financially by Research Institute for Astronomy \& Astrophysics of
Maragha (RIAAM), Iran.


\section*{Appendix: Holographic renormalization}

\bigskip To construct the boundary counterterm action, we utilize a
holographic renormalization method of Skenderis which was presented in \cite%
{Sken,Bar,Sun}. To apply this method, the spacetime metric takes the form%
\begin{equation}
ds^{2}=G_{ab}d\xi ^{a}d\xi ^{b}=-\frac{L_{\mathrm{eff}}^{2}}{\varrho }%
\mathcal{X}(\varrho )d\tau ^{2}+\frac{L_{\mathrm{eff}}^{2}d\varrho ^{2}}{%
4\varrho \mathcal{X}(\varrho )}+\frac{L_{\mathrm{eff}}^{2}}{\varrho }%
(dx^{2}+dy^{2}+dz^{2}),  \label{metapp}
\end{equation}%
which relates to the metric (\ref{metr}) via $\tau =t/L_{\mathrm{eff}}$, $%
\varrho =L_{\mathrm{eff}}^{2}/r$ and the asymptotic ($\varrho \rightarrow 0$%
) metric function is $\mathcal{X}(\varrho )_{\varrho \rightarrow 0}=1$.
Hence, asymptotically metric becomes%
\begin{equation}
ds^{2}=\frac{L_{\mathrm{eff}}^{2}d\varrho ^{2}}{4\varrho \mathcal{X}(\varrho
)}+h_{\mu \nu }dx^{\mu }dx^{\nu },
\end{equation}%
where%
\begin{equation}
h_{\mu \nu }=\frac{L_{\mathrm{eff}}^{2}}{\varrho }\gamma _{\mu \nu }^{0}.
\end{equation}%
According to the electromagnetic contribution, one can evaluate
the on-shell action as
\begin{eqnarray}
S &=&\int_{\mathcal{M}}d^{5}\xi \sqrt{-G}\left[ \mathcal{F}+b\mathcal{F}^{2}%
\right]   \notag \\
&=&\int_{\mathcal{M}}d^{5}\xi \frac{\sqrt{-G}}{2}A_{b}\nabla _{a}\left[
\left( 1+2b\mathcal{F}\right) F^{ab}\right] -\int_{\partial \mathcal{M}%
}d^{4}x\frac{\sqrt{-h}}{2}A_{\mu }F^{\varrho \mu }\left( 1+2b\mathcal{F}%
\right)   \notag \\
&=&-L_{\mathrm{eff}}\int_{\varrho =\epsilon }d^{4}x\sqrt{-\gamma ^{0}}A_{\mu
}\partial _{\varrho }A_{\nu }\gamma ^{0\mu \nu }\left( 1+2b\mathcal{F}%
\right) ,
\end{eqnarray}%
where $\epsilon $ is a small constant parameter. On the new coordinates (\ref%
{metapp}), the gauge field equation of bulk motion near boundary is given by
\begin{equation}
\varrho \frac{d^{2}A_{i}}{d\varrho ^{2}}+\frac{1}{4}\partial _{0}^{2}A_{i}+%
\mathcal{O}(\varrho ^{2})=0,
\end{equation}%
where $\partial _{0}^{2}$ points out to the wave operator of boundary metric
$\gamma _{\mu \nu }^{0}$ and the general solution of this equation is%
\begin{equation}
A_{i}=A_{i}^{0}+A_{i}^{1}\varrho +\psi \varrho \mathrm{ln}(\varrho ),
\label{ansA}
\end{equation}%
where $\psi =-1/4\partial _{0}^{2}A_{i}^{0}$. With $A_{i}$ at hand, the
on-shell electromagnetic action can be written as%
\begin{equation}
S=-L_{\mathrm{eff}}\int_{\varrho =\epsilon }d^{4}x\sqrt{-\gamma ^{0}}%
A_{i}^{0}\left( A_{i}^{1}-\frac{1}{4}\partial _{0}^{2}A_{i}^{0}-\frac{1}{4}%
\mathrm{ln}(\epsilon )\partial _{0}^{2}A_{i}^{0}\right) ,
\end{equation}%
which is logarithmically divergent. Now, according to Ref.\cite{Sken}, in
order to determine the counterterm action, we first invert the solution (\ref%
{ansA}) to give $A_{i}^{0}=A_{i}+\mathcal{O}(\epsilon )$. Thus, the
counterterm action is obtained as%
\begin{eqnarray}\label{SCt}
S_{c.t.} &=&-\frac{L_{\mathrm{eff}}}{4}\mathrm{ln}(\epsilon
)\int_{\varrho
=\epsilon }d^{4}x\sqrt{-\gamma ^{0}}A_{i}\partial _{0}^{2}A_{i}  \notag \\
&=&-\frac{L_{\mathrm{eff}}}{4}\mathrm{ln}(\epsilon )\int_{\varrho =\epsilon
}d^{4}x\frac{\sqrt{-h}}{2}F_{\mu \nu }F^{\mu \nu }.
\end{eqnarray}%
It is notable to mention that since we consider $A_{i}=A_{i}(\varrho )\exp
(-i\omega L_{\mathrm{eff}}\tau )$, one can calculate
\begin{equation}
\partial _{0}^{2}A_{i}=\omega ^{2}L_{\mathrm{eff}}^{2}A_{i}.
\end{equation}


\end{document}